\documentclass[12pt,twoside]{article}
\usepackage{correl13}

\def\TheTitle{Strongly correlated superconductivity}
\def\TheHeading{Strongly correlated superconductivity}
\def\TheAuthor{A.-M.S. Tremblay}
\def\TheInstitute{D\'epartement de physique \\ Regroupement qu\'eb\'ecois sur les mat\'eriaux de pointe 
\\ Canadian Institute for Advanced Research}
\def\TheAddress{Sherbrooke, Qu\'ebec J1K 2R1, Canada}
\def\TheChapter{10}                       

\begin{document}
\MakeTitle           
\section{Introduction}
	Band theory and the BCS-Eliashberg theory of superconductivity are arguably the most successful theories of condensed matter physics by the breadth and subtlety of phenomena they explain.  Experimental discoveries however clearly signal their failure in certain cases. Around 1940, it was discovered that some materials with an odd number of electrons per unit cell, for example NiO, were insulators instead of metals, a failure of band theory~\cite{Boer:1937}. Peierls and Mott quickly realized that strong effective repulsion between electrons could explain this (Mott) insulating behaviour.~\cite{Imada:1998} In 1979 and 1980, heavy fermion \cite{steglich_superconductivity_1979} and organic \cite{Jerome:1980} superconductors were discovered, an apparent failure of BCS theory because the proximity of the superconducting phases to antiferromagnetism suggested the presence of strong electron-electron repulsion, contrary to the expected phonon-mediated attraction that gives rise to superconductivity in BCS. Superconductivity in the cuprates~\cite{Bednorz:1986} in layered organic superconductors~\cite{Williams:1991,Jerome:1991} and in the pnictides~\cite{Hosono:2008} eventually followed the pattern: superconductivity appeared at the frontier of antiferromagnetism and, in the case of the layered organics, at the frontier of the Mott transition,~\cite{Ito:1996,Kanoda:1997} providing even more examples of superconductors falling outside the BCS paradigm.~\cite{Lefebvre:2000,LimeletteBEDT:2003} The materials that fall outside the range of applicability of band and of BCS theory are often called strongly correlated or quantum materials. They often exhibit spectacular properties, such as colossal magnetoresistance, giant thermopower, high-temperature superconductivity etc. 
	
	The failures of band theory and of the BCS-Eliashberg theory of superconductivity are in fact intimately related. In these lecture notes, we will be particularly concerned with the failure of BCS theory, and with the understanding of materials belonging to this category that we call strongly correlated superconductors. These superconductors have a normal state that is not a simple Fermi liquid and they exhibit surprising superconducting properties. For example, in the case of layered organic superconductors, they become better superconductors as the Mott transition to the insulating phase is approached.~\cite{Powell:2006}\footnote{See Fig. 6 of this review} 
	
	These lecture notes are not a review article. The field is still evolving rapidly, even after more than 30 years of research. My aim is to provide for the student at this school an overview of the context and of some important concepts and results. I try to provide entries to the literature even for topics that are not discussed in detail here. Nevertheless, the reference list is far from exhaustive. An exhaustive list of all the references for just a few sub-topics would take more than the total number of pages I am allowed.   
	
	I will begin by introducing the one-band Hubbard model as the simplest model that contains the physics of interest, in particular the Mott transition. That model is 50 years old this year,~\cite{Hubbard:1963,Kanamori:1963,Gutzwiller:1963} yet it is far from fully understood. Section~\ref{Sec:AFM} will use antiferromagnetism as an example to introduce notions of weak and strong correlations and to contrast the theoretical methods that are used in both limits. Section~\ref{Sec:Superconductivity} will do the same for superconductivity. Finally, Section~\ref{Sec:QuantumCluster} will explain some of the most recent results obtained with Cluster generalizations of Dynamical Mean-Field theory, approaches that allow one to explore the weak and strong correlation limits and the transition between both. 
	

\section{The model}\index{Hubbard model}\label{Hubbard model}
The one-band Hubbard model is given by
\begin{equation}
H=-\sum_{i,j,\sigma}t_{ij}c_{i\sigma}^{\dagger}c_{j\sigma}+U\sum
_{i}n_{i\uparrow}n_{i\downarrow}%
\label{Hubbard-model}
\end{equation}
where $i$ and $j$ label Wannier states on a lattice, $c_{i\sigma}^{\dagger}$ ($c_{i\sigma}$) are creation and annihilation
operators for electrons of spin $\sigma$, $n_{i\sigma}=c_{i\sigma}^{\dagger
}c_{i\sigma}$ is the density of spin $\sigma$ electrons, $t_{ij}=t_{ji}^{\ast
}$ is the hopping amplitude, that can be taken as real in our case, and $U$ is the on-site Coulomb repulsion. In general, we write $t,t^{\prime},t^{\prime\prime}$ respectively for the first-,
second- and third-nearest neighbour hopping amplitudes.

This is a drastic simplification of the complete many-body Hamiltonian, but we want to use it to understand the physics from the simplest point of view, without a large number of parameters. The first term of the Hubbard model Eq.~\eqref{Hubbard-model} is diagonal in a momentum-space single-particle basis. There, the wave nature of the electron is manifest. If the interaction $U$ is small compared to the bandwidth, perturbation theory and Fermi liquid theory hold.~\cite{nozieres_derivation_1962B,luttinger_derivation_1962B} This is the so-called weak-coupling limit. 

The interaction term in the Hubbard Hamiltonian, proportional to $U$, is diagonal in position space, i.e. in the Wannier orbital basis, exhibiting the particle nature of the electron. The motivation for that term is that once the interactions between electrons are screened, the dominant part of the interaction is on-site. Strong-coupling perturbation theory can be used if the bandwidth is small compared with the interaction~\cite{harris:1967,metzner_linked-cluster_1991,bartkowiak_magnetic_1992,Pairault:1998,Pairault:2000}. Clearly, the intermediate-coupling limit will be most difficult, the electron exhibiting both wave and particle properties at once. The ground state will be entangled, i.e. very far from a product state of either Bloch (plane waves) of Wannier (localized) orbitals. We refer to materials in the strong or intermediate-coupling limits as strongly correlated. 

When the interaction is the largest term and we are at half-filling, the solution of this Hamiltonian is a Mott insulating state. The ground state will be antiferromagnetic if there is not too much frustration. That can be seen as follows. If hopping vanishes, the ground state is $2^N$-fold degenerate if there are $N$ sites. Turning-on nearest-neighbour hopping, second order degenerate perturbation theory in $t$ leads to an antiferromagnetic interaction $J\mathbf{S_i}\cdot\mathbf{S_j}$ with $J=4t^2/U$.~\cite{gros_antiferromagnetic_1987,anderson_new_1959} This is the Heisenberg model. Since $J$ is positive, this term will be smallest for anti-parallel spins. The energy is generally lowered in second-order perturbation theory. Parallel spins cannot lower their energy through this mechanism because the Pauli principle forbids the virtual, doubly occupied state.  P.W. Anderson first proposed that the strong-coupling version of the Hubbard model could explain high-temperature superconductors.~\cite{Anderson:1987}

\begin{figure}[t!]
 \centering
 \includegraphics[width=1.0\textwidth]{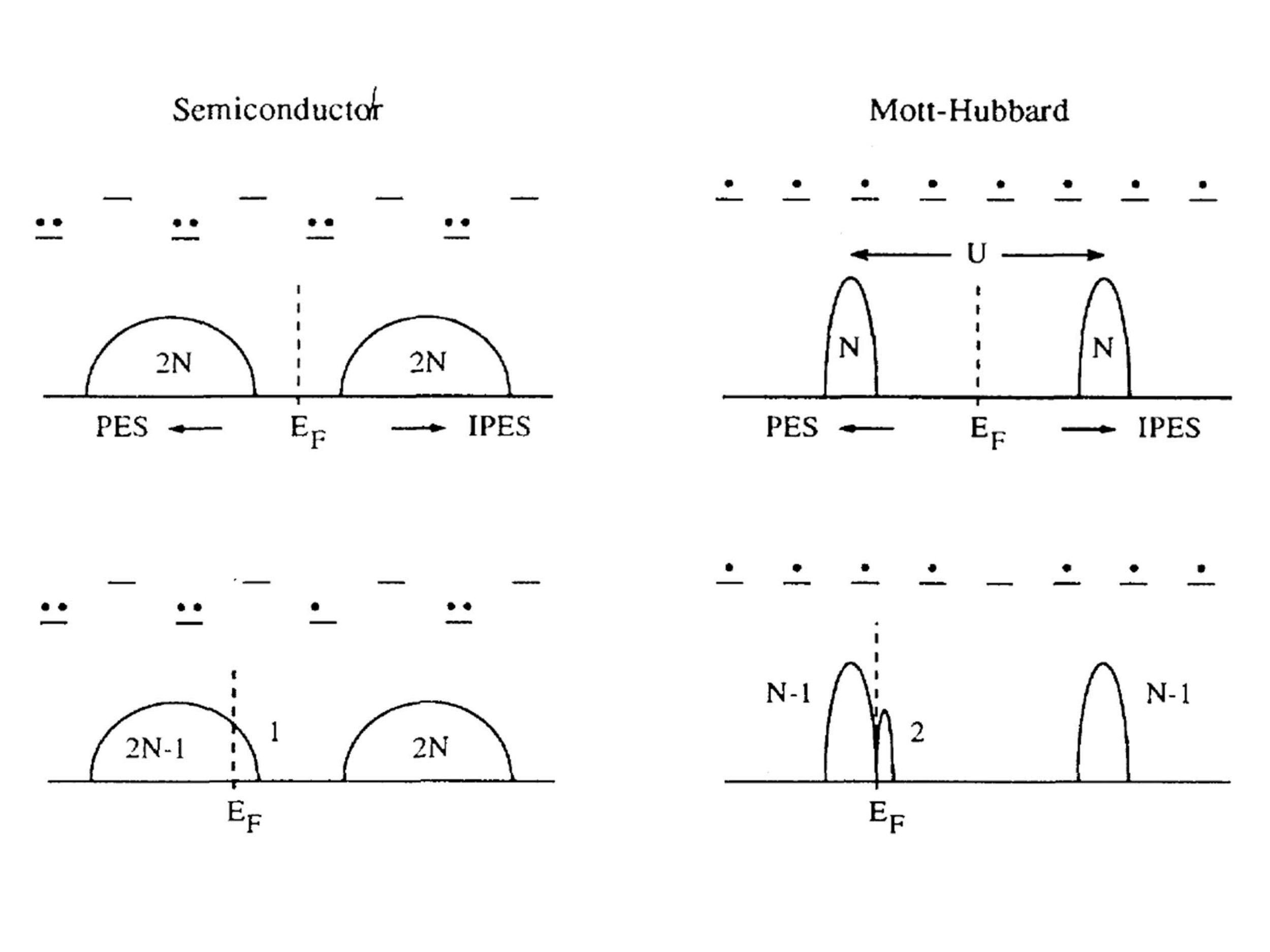}
 \caption{Figure from Ref.~\cite{meinders_spectral-weight_1993}. In a band insulator, as illustrated on the top-left figure, the valence band is filled. For $N$ sites on the lattice, there are $2N$ states in the valence band, the factor of $2$ accounting for spins, and $2N$ states in the conduction band. PES on the figures refers to ``Photoemission Spectrum'' and IPES to ``Inverse Photoemission Spectrum''. The small horizontal lines represent energy levels and the dots stand for electrons. In a Mott insulator, illustrated on the top-right figure, there are $N$ states in the lower energy band (Lower Hubbard band) and $N$ in the higher energy band (Upper Hubbard band), for a total of $2N$ as we expect in a single band. The two bands are separated by an energy $U$ because if we add an electron to the already occupied states, it costs an energy $U$. Perhaps the most striking difference between a band and a Mott insulator manifests itself when the Fermi energy $E_F$ is moved to dope the system with one hole. For the semiconductor, the Fermi energy moves, but the band does not rearrange itself. There is one unoccupied state right above the Fermi energy. This is seen on the bottom-left figure. On the bottom-right figure, we see that the situation is very different for a doped Mott insulator. With one electron missing, there are two states  just above the Fermi energy, not one state only. Indeed one can add an electron with a spin up or down on the now unoccupied site. And only $N-1$ states are left that will cost an additional energy $U$ if we add an electron. Similarly, $N-1$ states survive below the Fermi energy.}
 \label{Fig:Meinders}
\end{figure}

A caricature of the difference between an ordinary band insulator and a Mott insulator appears in Figure \ref{Fig:Meinders}. Refer to the explanations in the caption. 

\section{Weakly and strongly correlated antiferromagnets}\index{antiferromagnetism}\label{Sec:AFM}
\index{emergent properties} A phase of matter is characterized by very general ``emergent'' properties, i.e. properties that are qualitatively different from those of constituent atoms.~\cite{Anderson:1972,laughlin_different_2006,Pines:2010} For example, metals are shiny and they transport DC current. These are not properties of individual copper or gold atoms. It takes a finite amount of energy to excite these atoms from their ground state, so they cannot transport DC current. Also, their optical spectrum is made of discrete lines whereas metals reflect a continuous spectrum of light at low energy. In other words, the Fermi surface is an emergent property. Even in the presence of interactions, there is a jump in momentum occupation number which defines the Fermi surface. This is the Landau Fermi liquid.~\cite{nozieres_derivation_1962B,luttinger_derivation_1962B} Emergent properties appear at low energy, i.e. for excitation energies not far from the ground state. The same emergent properties arise from many different models. (In the renormalization group language, phases are trivial fixed points, and many Hamiltonians flow to the same fixed point). 

In this section, we use the antiferromagnetic phase to illustrate further was is meant by an emergent property and what properties of a phase depend qualitatively on whether we are dominated by band effects, or by strong correlations. Theoretical methods appropriate for each limit are described in the last subsection. 
    \subsection{Antiferromagnets: A qualitative discussion}\index{antiferromagnetism}\index{antiferromagnetism!weak coupling}\index{antiferromagnetism!strong coupling}
Consider the nearest-neighbour Hubbard model at half-filling on the cubic lattice in three dimensions. At $T=0$, there is a single phase, an antiferromagnet, whatever the value of the interaction $U$. One can increase $U$ continuously without encountering a phase transition. There is an order parameter in the sense of Landau, in this case the staggered magnetization. This order parameter reflects the presence of a broken symmetry: time reversal, spin rotational symmetry and translation by a lattice spacing are broken while time reversal accompanied by translation by a lattice spacing is preserved. 

A single-particle gap and spin waves as Goldstone modes are emergent consequences of this broken symmetry. Despite the fact that we are in a single phase, there are qualitative differences between weak and strong coupling as soon as we probe higher energies. For example, at strong-coupling spin waves persist all the way to the zone boundary and energy scale $J$ whereas at weak coupling spin waves enter the particle-hole continuum and become Landau damped before we reach the zone boundary. The ordered moment is saturated to its maximum value when $U$ is large enough but it can become arbitrarily small as $U$ decreases.

\begin{figure}[t!]
 \centering
 \includegraphics[width=1.0\textwidth]{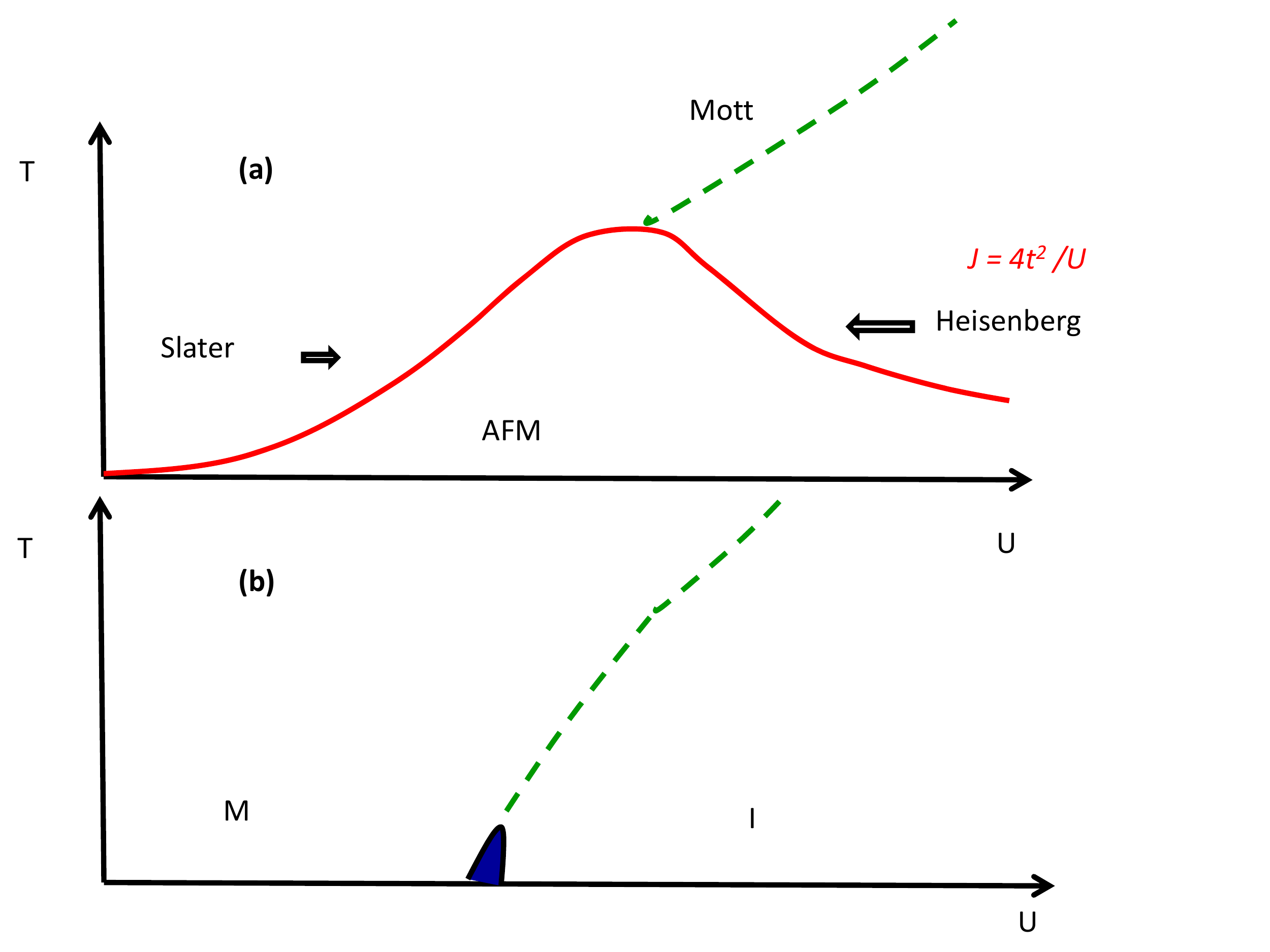}
 \caption{Schematic phase diagram of the 3d Hubbard model in the temperature $T$ interaction $U$ plane for perfect nesting. (a) The solid red line is the N\'eel temperature $T_N$ below which the system is antiferromagnetic. Coming from the left, it is a metallic state that becomes unstable to antiferromagnetism by the Slater mechanism. Coming from the right, it is a gapped insulator with local moments described by the Heisenberg model that becomes unstable to antiferromagnetism. The dashed green line above the maximum $T_N$ indicates a crossover from a metallic state with a Fermi surface to a gapped state with local moments. That crossover can be understood in (b) where antiferromagnetism is prevented from occurring. There dynamical mean-field theory predicts a first-order phase transition between a metal and a Mott insulator, with a coexistence region indicated in blue.}
 \label{Fig:AFM_phase_diagram}
\end{figure}

The differences between weak and strong coupling are also striking at finite temperature. This is illustrated in a schematic fashion in Fig.~\ref{Fig:AFM_phase_diagram}(a). The red line, so-called N\'eel temperature $T_N$, increases as we increase $U$ because the instability of the normal state fundamentally comes from nesting. In other words, thinking again of perturbation theory, the flat parts of the Fermi surface are connected by the antiferromagnetic wave vector $Q=(\pi,\pi)$ which means vanishing energy denominator at that wave vector and thus large spin susceptibility $\chi$ with a phase transition occuring when $U\chi$ is large enough. At strong coupling, $T_N$ decreases with increasing $U$ because the spin stiffness is proportional to $J=4t^2/U$.  Since we can in principle vary the ratio $t/U$ by changing pressure, it is clear that the pressure derivative of the N\'eel temperature has opposite sign at weak and strong coupling. The normal state is also very different. If we approach the transition from the left, as indicated by the arrow marked "Slater",\index{antiferromagnetism!Slater} we are in a metallic phase. We also say that the antiferromagnet that is born out of this metallic phase is an itinerant antiferromagnet.\index{antiferromagnetism!itinerant} On the contrary, approaching the transition from the right, we come from an insulating (gapped) phase described by the Heisenberg model. Increasing $U$ at fixed $T$ above the maximum $T_N$, we have to crossover from a metal, that has a Fermi surface, to an insulator, that has local moments, a gap and no Fermi surface. This highly non-trivial Physics is indicated by a dashed green line in Fig.~\ref{Fig:AFM_phase_diagram}(a) and marked ``Mott''.~\cite{Reymbaut:2013} 

\index{Mott transition}\index{Mott transition!crossover} Fig.~\ref{Fig:AFM_phase_diagram}(b) illustrates another way to understand the dashed green line. Imagine that antiferromagnetism does not occur before we reach zero temperature. This can be achieved in two dimensions where the Mermin-Wagner-Hohenberg~\cite{Mermin:1966,Hohenberg:1966} theorem prevents a broken continuous symmetry at finite temperature.\index{Mermin-Wagner-Hohenberg theorem} More generally antiferromagnetism can be prevented by frustration.~\cite{Toulouse:1977} Frustration can come either from longer-range hopping that lead to longer-range antiferromagnetic interactions or from the geometry of the lattice. (e.g. the triangular lattice). In either case, it becomes be impossible to minimize the energy of all the individual antiferromagnetic bonds without entering a contradiction.\index{frustration} In Dynamical Mean-Field theory,~\cite{Georges:1992,Jarrell:1992,Georges:1996} that we will discuss later, antiferromagnetism can just be prevented from occurring in the theory. In any case, what happens if antiferromagnetism is prevented is a first-order transition at $T=0$ between a metal and an insulator. This is the Mott phase transition that ends at a critical temperature. This transition is seen, for example, in layered organic superconductors of the $\kappa$-BEDT family that we will briefly discuss later.~\cite{Williams:1991,Jerome:1991} The dashed green line at finite temperature is the crossover due to this transition. In the case where the antiferromagnetic phase is artificially prevented from occurring, the metallic and insulating phases at low temperature are metastable phases in the same way that the normal metal is a metastable state below the superconducting transition temperature. Just as it is useful to think of a Fermi liquid at zero temperature even when it is not the true ground state, it is useful to think of the zero-temperature Mott insulator even when it is a metastable state.   
    
%
%
%
    
    \subsection{Contrasting methods for weak and strong coupling antiferromagnets and their normal state}
In this subsection, I list some of the approaches that can be used to study the various limiting cases as well as their domain of applicability wherever possible. Note that the antiferromagnetic state can occur away from half-filling as well, so we also discuss states that would be best characterized as Fermi liquids.
        \subsubsection{Ordered state} 
The ordered state at weak coupling can be described for example by mean-field theory applied directly to the Hubbard model.~\cite{Zhang:1989} Considering spin waves as collective modes, one can proceed by analogy with phonons and compute the corresponding self-energy resulting from the exchange of spin waves. The staggered moment can then be obtained from the resulting Green function. It seems that this scheme interpolates smoothly and correctly from weak to strong coupling. More specifically, at very strong coupling in the $T=0$ limit, the order parameter is renormalized down from its bare mean-field value by an amount very close to that predicted in a localized picture with a spin wave analysis of the Heisenberg model: In two dimensions on the square lattice, only two thirds of the full moment survives the zero-point fluctuations.~\cite{reger_monte_1988} This is observed experimentally in the parent high-temperature superconductor La$_2$CuO$_4$~\cite{vaknin_antiferromagnetism_1987}. 

At strong coupling when the normal state is gapped, one can perform degenerate perturbation theory, or systematically apply canonical transformations to obtain an effective model~\cite{takahashi_half-filled_1977,Macdonald:1988} that reduces to the Heisenberg model at very strong coupling. When the interaction is not strong enough, higher order corrections in $t/U$ enter in the form of longer-range exchange interactions and so-called ring-exchange.~\cite{takahashi_half-filled_1977,Macdonald:1988,delannoy:2005,yang_effective_2010,dalla_piazza_unified_2012} Various methods such as $1/S$ expansion~\cite{holstein_field_1940} 1/N expansion~\cite{chubukov_universal_1993} or non-linear sigma model~\cite{haldane_nonlinear_1983} are available. 

Numerically, stochastic series expansion,~\cite{Sandvik:2011} high-temperature series expansion,~\cite{Compostrini_high_temperature_series:2002} Quantum Monte Carlo (QMC)~\cite{kozik_netemperature_2013} world-line or worm algorithms~\cite{prokofev_worm_1998,boninsegni_worm_2006} and variational methods~\cite{sorella_wave_2005} are popular and accurate. Variational methods can be biased since one must guess a wave function. Nevertheless, variational methods~\cite{Giamarchi:1991} and QMC have shown that in two dimensions, on the square lattice, the antiferromagnetic ground state is the most likely ground state.~\cite{manousakis_spin-_1991} States described by singlet formation at various length scales, so-called Resonating Valence Bond spin liquids, are less stable. For introductions to various numerical methods, see the web archives of the following two summer schools.~\cite{ecole:2008,ecole:2012}

    
        \subsubsection{The normal state}\index{Two-Particle Self-Consistent theory (TPSC)} \index{RPA}\index{SCR}\index{FLEX}\label{MethodsNormalState}
In strong coupling, the normal state is an insulator described mostly by the non-linear sigma model.~\cite{chakravarty_two-dimensional_1989,keimer_magnetic_1992} In the weak-coupling limit, the normal state is a metal described by Fermi liquid theory. To describe a normal state that can contain strong antiferromagnetic fluctuations,~\cite{monthoux_yba_2cu_3o_7:_1993} one needs to consider a version of Fermi liquid theory that holds on a lattice. Spin propagates in a diffusive manner. These collective modes are known as paramagnons. The instability of the normal state to antiferromagnetism can be studied by the Random Phase Approximation (RPA), by Self-Consistent-Renormalized theory (SCR)~\cite{Moriya:1985} by the Fluctuation Exchange Approximation~\cite{Bickers_dwave:1989,Bickers:1989}, by the Functional Renormalization Group (FRG)~\cite{ZanchiSchulz:2000,HonerkampFRG:2001,shankar_renormalization-group_1994}, by field-theory methods~\cite{kaul_destruction_2008,sachdev_fluctuating_2009}, and by the Two-Particle Self-Consistent Approach (TPSC),~\cite{Vilk:1997,Allen:2003,TremblayMancini:2011} to give some examples. Numerically, Quantum Monte Carlo (QMC) is accurate and can serve as benchmark, but in many cases it cannot go to very low temperature because of the sign problem. That problem does not occur in the half-filled nearest-neighbor one-band Hubbard model, which can be studied at very low temperature with QMC.~\cite{bulut_electronic_1994} 

The limitations of most of the above approaches have been discussed in Appendices of Ref.~\cite{Vilk:1997}. Concerning TPSC, in short it is non-perturbative and is the most accurate of the analytical approaches at weak to intermadiate coupling, as judged from benchmark QMC~\cite{Vilk:1997,LTP:2006,TremblayMancini:2011}. TPSC also satisfies the Pauli principle in the sense that the square of the occupation number for one spin species on a lattice site is equal to the occupation number itself, in other words $1^2=1$ and $0^2=0$. RPA is an example of a well known theory that violates this constraint.\footnote{Appendix A3 of Ref.~\cite{Vilk:1997}} Also, TPSC satisfies conservation laws and a number of sum rules, including those which relate the spin susceptibility to the local moment and the charge susceptibility to the local charge. Most importantly, TPSC satisfies the Mermin-Wagner-Hohenberg theorem in two dimensions, contrary to RPA. Another effect included in TPSC is the renormalization of $U$ coming from cross channels (Kanamori, Brückner screening).~\cite{Brueckner:1960,Kanamori:1963} On a more technical level, TPSC does not assume a Migdal theorem in the calculation of the self-energy and the trace of the matrix product of the self-energy with the Green function satisfies the constraint that it is equal to twice the potential energy. 
        
The most important prediction that came out of TPSC for the normal state on the two-dimensional square lattice, is that precursors of the antiferromagnetic ground state will occur when the antiferromagnetic correlation length becomes larger than the thermal de Broglie wave length $\hbar v_F/k_B T$.~\cite{Vilk:1995,Vilk:1996,Moukouri:2000} The latter length is defined by the inverse of the wave vector spread that is caused by the thermal excitation $\Delta\varepsilon\sim k_BT$. This result was verified experimentally~\cite{Motoyama:2007} and it explains the pseudogap in electron-doped high-temperature superconductors.~\cite{Kyung:2004,TremblayMancini:2011,Bergeron:2012} 

\section{Weakly and strongly correlated superconductivity}\index{superconductivity}\label{Sec:Superconductivity}
As discussed in the previous section, antiferromagnets have different properties depending on whether $U$ is above or below the Mott transition and appropriate theoretical methods must be chosen depending on the case. In this section, we discuss the analogous phenomenon for superconductivity. A priori, the superconducting state of a doped Mott insulator or of a doped itinerant antiferromagnet are qualitatively different, even though some emergent properties are similar. 
    \subsection{Superconductors: A qualitative discussion}\label{SuperconductorsQualitative}
As for antiferromagnets, the superconducting phase has emergent properties. For an s-wave superconductor, global charge conservation, or $U(1)$ symmetry, is broken. For a d-wave superconductor, in addition to breaking $U(1)$ symmetry, the order parameter does not transform trivially under rotation by $\pi/2$. It breaks C$_{4v}$ symmetry on the square lattice. In both cases we have singlet superconductivity so spin-rotational symmetry is preserved. In both cases, long-range forces push the Goldstone modes to the plasma frequency by the Anderson-Higgs mechanism.~\cite{Varma:2002} 
The presence of symmetry-dictated nodes in the d-wave case is an emergent property with important experimental consequences: for example, the specific heat will vanish linearly with temperature as $T$ vanishes, and similarly for $\kappa$ the thermal conductivity. The ratio $\kappa/T$, reaches a universal constant in the $T=0$ limit, i.e. that ratio is independent of disorder.~\cite{lee_localized_1993,taillefer_universal_1997} The existence of a single-particle gap with nodes determined by symmetry is also an emergent property, but its detailed angular dependence and its size relative to other quantities, such as the transition temperature $T_c$ is dependent on details. 

\index{superconductivity!Eliashberg} A possible source of confusion in terminology is that in the context of phonon mediated s-wave superconductivity, there is the notion of strong-coupling superconductivity. The word strong-coupling has a slightly different meaning from the one discussed up to now. The context should make it clear what we are discussing. Eliashberg theory describes phonon-mediated strong-coupling superconductivity.~\cite{Eliashberg:1960,Carbotte:1990,Marsiglio:2001,Schrieffer:1964,Scalapino:1966} 
In that case, quasiparticles survive the strong electron-phonon interaction, contrary to the case where strong electron-electron interaction destroy the quasiparticles in favour of local moments in the Mott insulator. 

There are important quantitative differences between BCS and Eliashberg superconductors. In the latter case, the self-energy becomes frequency dependent so one can measure the effect of phonons on a frequency dependent gap function that influences in turn the tunnelling spectra. Predictions for the critical field $H_c(T)$ or for the ratio of the gap to $T_c$ for example differ. The Eliashberg approach is the most accurate.  

\index{superconductivity!weakly correlated}\index{superconductivity!strongly correlated} Let us return to our case, namely superconductors that arise from a doped Mott insulator (strongly correlated) and superconductors that arise from doping an itinerant antiferromagnet (weakly correlated). Again there are differences between both types of superconductors when we study more than just asymptotically small frequencies. For example, as we will see in Section~\ref{Sec:QuantumCluster}, in strongly correlated superconductors, the gap is no-longer particle-hole symmetric, and the transition temperature sometimes does not scale like the order parameter. A strongly-coupled superconductor is also more resilient to nearest-neighbor repulsion than a weakly-coupled one.~\cite{SenechalResilience:2013}

We should add a third category of strongly-correlated superconductors, namely superconductors that arise, in the context of the Hubbard model, at half-filling under a change of pressure. This is the case of the layered organics. There again superconductivity is very special, since contrary to naive expectations, it becomes stronger as we approach the Mott metal-insulator transition.~\cite{Powell:2006} 

Just as for antiferromagnets, the normal state of weakly-correlated and of strongly-correlated superconductors is very different. Within the one-band Hubbard model as usual, the normal state of weakly-correlated superconductors is a Fermi liquid with antiferromagnetic fluctuations. In the case of strongly correlated superconductors, the normal state exhibits many strange properties, the most famous of which is probably the linear temperature dependence of resistivity that persists well above the Mott-Ioffe-Regel (MIR) limit.\index{Mott-Ioffe-Regel limit}~\cite{hussey__universality_2004,gurvitch_resistivity_1987} This limit is defined as follows. Consider a simple Drude formula for the conductivity, $\sigma=ne^2\tau/m$, where $n$ is the density, $e$ is the electron charge, $m$ the mass and $\tau$ the collision time. Using the Fermi velocity $v_F$, one can convert the scattering time $\tau$ to a mean-free path $\ell$. The MIR minimal conductivity is determined by stating that the mean-free path cannot be smaller than the Fermi wavelength. This means that, as a function of temperature, resistivity should saturate to that limit. This is seen in detailed many-body calculations with TPSC~\cite{Bergeron:2011} (These calculations do not include the possibility of the Mott transition). For a set of two-dimensional planes separated by a distance $d$, the MIR limit is set by $\hbar d/e^2$. That limit can be exceeded in doped Mott insulators.~\cite{XuKotliar:2013}

The strange-metal properties, including the linear temperature dependence of the resistivity, are often considered as emergent properties of a new phase. Since new phases of matter are difficult to predict, one approach has been to find mean-field solutions of gauge-field theories. These gauge theories can be derived from Hubbard-Stratonovich transofrmations or from assumptions as to the nature of the emergent degrees of freedom.~\cite{LeeRMP:2006} 

\begin{figure}[t!]
 \centering
 \includegraphics[width=0.7\textwidth]{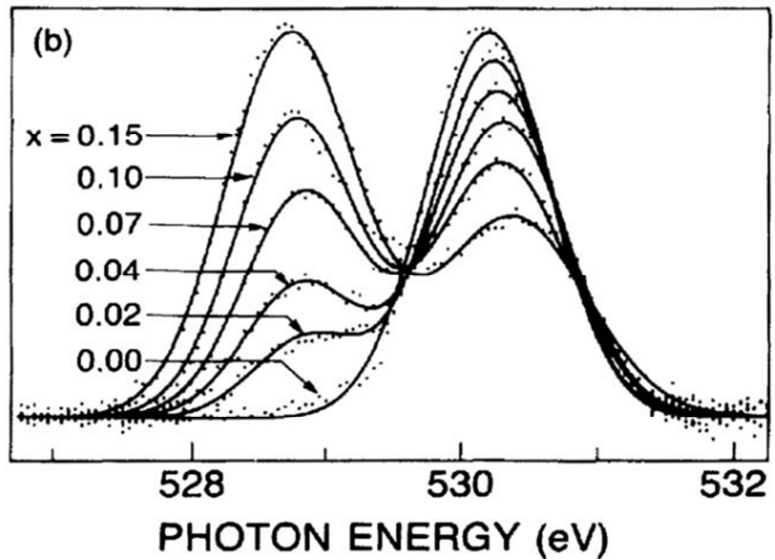}
\caption{This is the absorption spectrum for a sharp atomic level whose energy is about 0.5keV below the Fermi level. At half-filling, or zero doping $\delta=0$, it takes about 530eV to excite an electron from the deep level to the upper Hubbard band. As one dopes, the upper Hubbard band still absorbs, but new states appear just above the Fermi level, allowing the X-ray to be absorbed at an energy about 2eV below the upper Hubbard band. These states are illustrated on the bottom right panel of Fig.~\ref{Fig:Meinders}. The important point, is that the spectral weight in the states just above the Fermi energy grows about twice as fast as the spectral weight in the upper Hubbard band decreases, in agreement with the cartoon picture of a doped Mott insulator. Figure from Ref.~\cite{ChenX:1991}}
 \label{Fig:XAS}
\end{figure}

As in the case of the antiferromagnet, the state above the optimal transition temperature in strongly correlated superconductors is a state where crossovers occur. There is much evidence that hole-doped high-temperature superconductors are doped Mott insulators. The high-temperature thermopower~\cite{obertelli_systematics_1992,honma_unified_2008} and Hall coefficient~\cite{ando_evolution_2004} are examples of properties that are those expected from Mott insulators. We can verify the doped Mott insulator nature of the hole-doped cuprates from the experimental results for soft- X-ray absorption spectroscopy as illustrated in Fig.~\ref{Fig:XAS}. This figure should be compared with the cartoon in Fig.~\ref{Fig:Meinders} above. For further details, refer to the caption of Fig.~\ref{Fig:XAS}. More recent experimental results on this topic~\cite{peets:2009} and comparison with the Hubbard model~\cite{PhillipsComment:2010} are available in the literature.

%
%
%

    \subsection{Contrasting methods for weakly and strongly correlated superconductors}
    In most phase transitions, a simple analysis at weak coupling indicates that the normal state is unstable towards a new phase. For example, one can compute an appropriate susceptibility in the normal state and observe that it can sometimes diverge at sufficiently low temperature, indicating an instability. For an antiferromagnet, we would compute the staggered spin-susceptibility. Alternatively, one could perform a mean-field factorization and verify that there is a self-consistent broken symmetry solution at low temperature. Neither of these two procedures however leads to a superconducting instability when we start from a purely repulsive Hubbard model. In this section, we will discuss how to overcome this.
    
    Surprisingly, a mean-field factorization of the Hubbard model at strong coupling does lead to a superconducting d-wave ground state.~\cite{Miyake:1986,Kotliar:1988,AndersonVanilla:2004} There are reasons however to doubt the approximations involved in a simple mean-field theory at strong coupling.   
    
        \subsubsection{The normal state and its superconducting instability}
        
        In the weakly correlated case, the normal state must contain slow modes that replace the phonons to obtain a superconducting instability. It is as if the superconducting instability came at a second level of refinement of the theory. This all started with Kohn and Luttinger~\cite{kohn:1965,Mraz:2003}: They noted that the interaction between two electrons is screened by the other electrons. Computing this screening to leading order, they found that in sufficiently high angular-momentum states and at sufficiently low temperature (usually very low) there is always a superconducting instability in a Fermi liquid.~\cite{maiti_superconductivity_2013}\index{Kohn-Luttinger theorem}\footnote{That reference contains a review.} It is possible to do calculations in this spirit that are exact for infinitesimally small repulsive interactions.~\cite{Raghu:2010} 
        
	Around 1986, before the discovery of high-temperature superconductivity, it was realized that if instead of a Fermi liquid in a continuum, one considers electrons on a lattice interacting with short-range repulsion, then the superconducting instability may occur more readily. More specifically, it was found that exchange of antiferromagnetic paramagnons between electrons can lead to a d-wave superconducting instability.~\cite{Beal-Monod:1986,Scalapino:1986} This is somewhat analogous to what happens for ferromagnetic fluctuations in superfluide $^3$He.~\footnote{For a recent exhaustive review of unconventional superconductivity, see Ref.~\cite{Norman:2000}} These types of theories,~\cite{Moriya:2003} like TPSC below, have some features that are qualitatively different from the BCS prediction.~\cite{monthoux_superconductivity_2007,ScalapinoThread:2010} For example, the pairing symmetry depends more on the shape of the Fermi surface than on the single-particle density of states. Indeed, the shape of the Fermi surface determines the wave vector of the largest spin fluctuations, which in turn favor a given symmetry of the d-wave order parameter, for example $d_{x^2-y^2}$ vs $d_{xy}$~\cite{Hassan:2008} depending on whether the antiferromagnetic fluctuations are in the $(\pi,\pi)$ or $(\pi,0)$ direction respectively.   It is also believed that conditions for maximal $T_c$ are realized close to the quantum critical point of an apparently competing phase, such as an antiferromagnetic phase,~\cite{broun_what_2008,SachdevQuantumCritical:2010} or a charge ordered phase~\cite{DoironTailleferQCP:2012}.

	In the case of quasi-one-dimensional conductors, a more satisfying way of obtaining d-wave superconductivity in the presence of repulsion consists in using the renormalization group.~\cite{Caron:1986,Sedeki:2012} In this approach, one finds a succession of effective low-energy theories by eliminating perturbatively states that are far away from the Fermi surface. As more and more degrees of freedom are eliminated, i.e. as the cutoff decreases, the effective interactions for the low-energy theory can grow or decrease, or even change sign. In this approach then, all fluctuation channels are considered simultaneously, interfering and influencing each other. The effective interaction in the particle-particle d-wave channel can become attractive, signalling a superconducting instability.    
	
	Whereas this approach is well justified in the one-dimensional and quasi one-dimensional cases from the logarithmic behavior of perturbation theory, in two dimensions more work is needed. Nevertheless, allowing the renormalized interactions to depend on all possible momenta, one can devise the so-called functional renormalization group. One can follow either the Wilson procedure~\cite{shankar_renormalization-group_1994},\index{functional renormalization group} as was done orginally for fermions by Bourbonnais,~\cite{Bourbonnais:1985,BourbonnaisHouches:1991} or a functional approach~\cite{ZanchiSchulz:2000,HonerkampFRG:2001} closer in spirit to quantum field-theory approaches. d-wave superconductivity has been found in these approaches.~\cite{HalbothPom:2000,Salmhofer:2004,MetznerReview:2012}    

	As we have already mentioned, in two dimensions, even at weak to intermediate coupling, the normal state out of which d-wave superconductivity emerges is not necessarily a simple Fermi liquid. It can have a pseudogap induced by antiferromagnetic fluctuations. Because TPSC is the only approach that can produce a pseudogap in two dimensions over a broad range of temperature, we mention some of the results of this approach.~\cite{Kyung:2003,Hassan:2008} In spirit, the approach is similar to the paramagnon theories above,~\cite{Beal-Monod:1986,Scalapino:1986,Moriya:2003} but it is more appropriate because, as mentioned before, it satisfies the Mermin-Wagner theorem, is non-perturbative and satisfies a number of sum rules. The irreducible vertex in the particle-particle channel is obtained from functional derivatives. Here are some of the main results:
	
	a) The pseudogap appears when the antiferromagnetic correlation length exceeds the thermal de Broglie wave length. This is why, even without competing order, $T_c$ decreases as one approaches half-filling despite the fact that the AFM correlation length increases. This can be seen in Fig. 3 of Ref~\cite{Kyung:2003}. States are removed from the Fermi level and hence they cannot lead to pairing. The dome is less pronounced when second-neighbor hopping $t^\prime$ is finite because the fraction of the Fermi surface where states are removed (hot spots) is smaller. 
	
	b) The superconducting $T_c$ depends rather strongly on $t^\prime$. At fixed filling, there is an optimal frustration, namely a value of $t^\prime$ where $T_c$ is maximum as illustrated in Fig. 5 of Ref.~\cite{Hassan:2008}. 
	
	c) Fig. 6 of Ref.~\cite{Hassan:2008} shows that $T_c$ can occur below the pseudogap temperature or above. (The caption should read $U=6$ instead of $U=4$). For the cases considered, that include optimal frustration, the AFM correlation length at the maximum $T_c$ is about 9 lattice spacings, as in Ref.~\cite{Varma:2013}. Elsewhere, it takes larger values at $T_c$. 
	
	d) A correlation between resistivity and $T_c$ in the pnictides, the electron-doped cuprates and the quasi one-dimensional organics was well established experimentally in Ref.~\cite{Doiron:2009}. Theoretically it is well understood for the quasi one-dimensional organics~\cite{Sedeki:2012}. \footnote{It seems to be satisfied in TPSC, as illustrated by Fig. 5 of Ref.~\cite{Bergeron:2012}, but the analytical continuation has some uncertainties.}    

	e) There is strong evidence that the electron-doped cuprates provide an example where antiferromagnetically mediated superconductivity can be verified. Fig. 3 and Fig. 4 of the paper by the Greven group~\cite{Motoyama:2007} show that at optimal $T_c$ the AFM correlation length is of the order of 10 lattice spacings. The photoemission spectrum and the AFM correlation length obtained from the Hubbard model~\cite{Kyung:2004} with $t^\prime=-0.175t, t''=0.05t$ and $U=6.25t$ agree with experiment. In particular, in TPSC one obtains the dynamical exponent $z=1$ at the antiferromagnetic quantum critical point,~\cite{Bergeron:2012} as observed in the experiments~\cite{Motoyama:2007}. This comes from the fact that at that quantum critical point, the Fermi surface touches only one point when it crosses the antiferromagnetic zone boundary. The strange discontinuous doping dependence of the AFM correlation length near the optimal $T_c$ obtained by Greven's group is however unexplained. The important interference between antiferromagnetism and d-wave superconductivity found with the functional renormalization group~\cite{Sedeki:2012} is not included in TPSC calculations however. I find that in the cuprate family, the electron-doped systems are those for which the case for a quantum-critical scenario~\cite{broun_what_2008,SachdevQuantumCritical:2010,DemlerSachdev:2002,YangJarrellZaanen:2011} for superconductivity is the most justified. 	

	In a doped Mott insulator, the problem becomes very difficult. The normal state is expected to be very anomalous, as we have discussed above. Based on the idea of emergent behavior, many researchers have considered slave-particle approaches.~\cite{LeeRMP:2006}  The exact creation-operation operators in these approaches are represented in a larger Hilbert space by products of fermionic and bosonic degrees of freedom with constraints that restrict the theory to the original Hilbert space. There is large variety of these approaches: Slave bosons of various kinds~\cite{Barnes:1976,Coleman:1984,KotliarRuckenstein:1986}, slave fermions,~\cite{Yoshida:1989,Jayaprakash:1989} slave rotors,~\cite{Florens:2002} slave spins,~\cite{HassanSlave:2010} or other field theory approaches.~\cite{PhillipsRMP:2010} Depending on which of these methods is used, one obtains a different kind of mean-field theory with gauge fields that are used to enforce relaxed versions of the exact constraints that the theories should satisfy. Since there are many possible mean-field theories for the same starting Hamiltonian that give different answers and no variational principle to decide between them,~\cite{Boies:1995} one must rely on intuition and on a strong belief on emergence in this kind of approaches.\index{slave particles}
	
	At strong coupling, the pseudogap in the normal state can also be treated phenomenologically quite successfully with the Yang Rice Zhang (YRZ) model.~\cite{Yang:2006} Inspired by renormalized mean-field theory, that I discuss briefly in the following section, this approach suggests a model for the Green function that can then be used to compute many observable properties.~\cite{SchachingerCarbotte:2010,LeBlancCarbotte:2011}
	
	The normal state may also be treated by numerical methods, such as variational approaches, or by quantum cluster approaches. Section \ref{Sec:QuantumCluster} below is devoted to this methodology.
	
	It should be pointed out that near the Mott transition at $U=6$ on the square lattice where TPSC ceases to be valid, the value of optimal $T_c$ that is found Ref.~\cite{Kyung:2003} is close to that found with the quantum cluster approaches discussed in Sec.\ref{Sec:QuantumCluster} below.~\cite{SordiSemon:2013} The same statement is valid at $U=4$~\cite{maier_d:2005,Kyung:2003,LTP:2006} where this time the quantum cluster approaches are less accurate than at larger $U$. This agreement of non-perturbative weak and strong coupling methods at intermediate coupling gives us confidence in the validity of the results. 
		         
%

        \subsubsection{Ordered state}

        Whereas the Hubbard model does not have a simple mean-field d-wave solution, its strong-coupling version, namely the t-J model does.~\cite{Miyake:1986,Kotliar:1988,AndersonVanilla:2004} More specifically if we perform second order degenerate perturbation theory starting from the large $U$ limit, the effective low-energy Hamiltonian reduces to
\begin{equation}
H=-\sum_{i,j,\sigma }t_{ij}Pc_{i\sigma }^{\dagger }c_{j\sigma
}P+J\sum_{\left\langle i,j\right\rangle }\mathbf{S}_{i}\cdot \mathbf{S}_{j}
\label{t-J-Model}
\end{equation}%
where $P$ are projection operators that ensure that hopping does not lead to double occupancy. In the above expression, correlated hopping terms and density-density terms have been neglected. 

To find superconductivity, one proceeds like Anderson~\cite{Anderson:1987} and writes the spin operators in terms of Pauli matrices $\vec{\sigma}$ and creation-annihilation operators so that the Hamiltonian reduces to
\begin{equation}
H=-\sum_{i,j,\sigma }t_{ij}Pc_{i\sigma }^{\dagger }c_{j\sigma
}P+J\sum_{\left\langle i,j\right\rangle, \alpha,\beta,\gamma,\delta }\left( \frac{1}{2}c_{i\alpha
}^{\dagger }\vec{\sigma }_{\alpha \beta }c_{i\beta }\right) \cdot \left( 
\frac{1}{2}c_{j\gamma }^{\dagger }\vec{\sigma }_{\gamma \delta }c_{j\delta
}\right). 
\end{equation}%
Defining the d-wave order parameter as follows, with $N$ the number of sites, and lattice spacing unity 
\begin{equation}
d=\left\langle \hat{d}\right\rangle =\frac{1}{N}\sum\limits_{\mathbf{k}%
}\left( \cos k_{x}-\cos k_{y}\right) \left\langle c_{\mathbf{k}\uparrow }c_{-%
\mathbf{k}\downarrow }\right\rangle ,
\end{equation}%
a mean-field factorization, including the possibility of N\'eel order $m$ leads to the mean-field Hamiltonian
\begin{equation}
H_{MF}=\sum\limits_{\mathbf{k,}\sigma }\varepsilon \left( \mathbf{k}\right)
c_{\mathbf{k}\sigma }^{\dagger }c_{\mathbf{k}\sigma }-4Jm\hat{m}%
-Jd\left( \hat{d}+\hat{d}^{\dagger }\right).
\label{H_MF}
\end{equation}
The dispersion relation $\varepsilon \left( \mathbf{k}\right)$ is obtained by replacing the projection operators by the average doping. The d-wave nature of the order was suggested in Refs.~\cite{Kotliar:1988,Inui:1988} The superconducting state in this approach is not much different from an ordinary BCS superconductor, but with renormalized hopping parameters. In the above approach, it is clear the instantaneous interaction $J$ causes the binding. This has led Anderson to doubt the existence of a ``pairing glue'' in strongly correlated superconductors.~\cite{Anderson:2007} We will see in the following section that more detailed numerical calculations give a different perspective.~\cite{Maier:2008,Kyung:2009} 

The intuitive weak-coupling argument for the existence of a d-wave superconductor in the presence of antiferromagnetic fluctuations~\cite{Scalapino:1995,ScalapinoThread:2010} starts from the BCS gap equation
\begin{equation}
\Delta _{\mathbf{p}}=-\int \frac{d\mathbf{p}^{\prime }}{(2\pi )^{2}}U\left( 
\mathbf{p-p}^{\prime }\right) \frac{\Delta _{\mathbf{p}^{\prime }}}{2E_{%
\mathbf{p}^{\prime }}}\left( 1-2f\left( E_{\mathbf{p}^{\prime }}\right)
\right) 
\end{equation}
where $E_{\mathbf{p}}=\sqrt{\varepsilon_\mathbf{p}^2+\Delta_\mathbf{p}^2}$ and $f$ is the Fermi function. In the case of an s-wave superconductor, the gap is independent of $\mathbf{p}$ so it can be simplified on both sides of the equation. There will be a solution only if $U$ is negative since all other factors on the right-hand side are positive. In the presence of a repulsive interaction, in other words when $U$ is positive, a solution where the order parameter changes sign is possible. For example, suppose $\mathbf{p}$ on the left-hand side is in the $(\pi,0)$ direction of a square lattice. Then if, because of antiferromagnetic fluctuations, $U\left(\mathbf{p-p}^{\prime }\right)$ is peaked near $(\pi,\pi)$, then the most important contributions to the integral come from points such that $\mathbf{p}^\prime$ is near $(0,\pi)$ or $(\pi,0)$, where the gap has a different sign. That sign will simplify with the overall minus sign on the right-hand side, making a solution possible. 

Superconductivity has also been studied with many strong-coupling methods, including the slave-particle-gauge-theory approaches~\cite{LeeRMP:2006}, the Composite-Operator Method~\cite{ManciniCOM:2011} and the YRZ approach mentioned above.~\cite{Yang:2006} In the next section, we focus on Quantum Cluster Approaches. 
        

\section{High-temperature superconductors and organics: the view from Dynamical Mean Field Theory}\label{Sec:QuantumCluster}

In the presence of a Mott transition, the unbiased numerical method of choice is dynamical mean-field theory. When generalized to a cluster,~\cite{Hettler:1998,Kotliar:2001,Lichtenstein:2000} one sometimes refers to these methods as ``Quantum Cluster approaches''. For reviews, see Refs.~\cite{Maier:2005,KotliarRMP:2006,LTP:2006}  The advantage of this method is that all short-range dynamical and spatial correlations are included. Long range spatial correlations on the other hand are included at the mean-field level as broken symmetry states. The symmetry is broken in the bath only, not on the cluster. Long-wavelength particle-hole and particle-particle fluctuations are, however, missing. 

After a short formal derivation of the method, the last two subsections will present a few results for the normal state, and for the superconducting state respectively. In both cases, we will emphasize the new physics that arises in the strong coupling regime.  

    \subsection{Quantum cluster approaches}
    
In short, Dynamical Mean-Field Theory (DMFT) can be understood simply as follows: In infinite dimension one can show that the self-energy depends only on frequency.~\cite{Metzner:1989} To solve the problem exactly, one considers a single site with a Hubbard interaction immersed in a bath of non-interacting electrons.~\cite{Georges:1992,Jarrell:1992,Georges:1996} Solving this problem, one obtains a self-energy that should be that entering the full lattice Green function. The bath is determined self-consistently by requiring that when the lattice Green function is projected on a single site, one obtains the same Green function as that of the single-site in a bath problem. In practice, this approach works well in three dimensions. In lower dimension, the self-energy acquires a momentum dependence and one must immerse a small interacting cluster in a self-consistent bath. One usually refers to the cluster, or the single-site as ``the impurity''.  The rest of this subsection is adapted from Ref.~\cite{LTP:2006}. It is not necessary to understand the details of this derivation to follow the rest of the lecture notes.  

Formally, the self-energy functional approach, devised by Potthoff\cite{Potthoff:2003,Potthoff:2003a,Potthoff:2003b,Potthoff:2011}
allows one to consider various cluster schemes from a unified point of view.
It begins with $\Omega_{\mathbf{t}}[G],$ a functional of the Green function%
\begin{equation}
\Omega_{\mathbf{t}}[G]=\Phi[G]-\mathrm{Tr}((G_{0\mathbf{t}}^{-1}%
-G^{-1})G)+\mathrm{Tr}\ln(-G). \label{GrandPotential}%
\end{equation}
The Luttinger Ward functional $\Phi[G]$ entering this equation is the
sum of two-particle irreducible skeleton diagrams. For our purposes,
what is important is that (1) The functional derivative of $\Phi[G]$ is
the self-energy%
\begin{equation}
\frac{\delta\Phi[G]}{\delta G}=\Sigma\label{SelfLuttinger}%
\end{equation}
and (2) it is a universal functional of $G$ in the following sense: whatever
the form of the one-body Hamiltonian, it depends only on the interaction and,
functionnally, it depends only on $G$ and on the interaction, not on the one-body Hamiltonian. The dependence of the functional $\Omega_{\mathbf{t}}[G]$ on the one-body part of the Hamiltonian is denoted by the subscript
$\mathbf{t}$ and it comes only through $G_{0\mathbf{t}}^{-1}$ appearing on the
right-hand side of Eq.~(\ref{GrandPotential}).

The functional $\Omega_{\mathbf{t}}[G]$ has the important property that it is
stationary when $G$ takes the value prescribed by Dyson's equation. Indeed,
given the last two equations, the Euler equation takes the form%
\begin{equation}
\frac{\delta\Omega_{\mathbf{t}}[G]}{\delta G}=\Sigma-G_{0\mathbf{t}}%
^{-1}+G^{-1}=0.
\end{equation}
This is a dynamic variational principle since it involves the frequency
appearing in the Green function, in other words excited states are involved in
the variation. At this stationary point, and only there, $\Omega_{\mathbf{t}%
}[G]$ is equal to the grand potential. Contrary to Ritz's variational
principle, this last equation does not tell us whether $\Omega_{\mathbf{t}%
}[G]$ is a minimum or a maximum or a saddle point at the extremum.

Suppose we can locally invert Eq.~(\ref{SelfLuttinger}) for the self-energy to write $G$ as a functional of $\Sigma.$ We can use this result to write,

\begin{equation}
\Omega_{\mathbf{t}}[\Sigma]=F[\Sigma]-\mathrm{Tr}\ln(-G_{0\mathbf{t}}^{-1}+\Sigma),
\end{equation}

where we defined

\begin{equation}
F[\Sigma]=\Phi[G]-\mathrm{Tr}(\Sigma G)
\end{equation}

and where it is implicit that $G=G[\Sigma]$ is now a functional of $\Sigma$. We refer to this functional as the Potthoff functional\index{Potthoff functional}. Potthoff called this method the self-energy functional approach. Several types of quantum cluster approaches may be derived from this functional. A crucial observation is that $F[\Sigma],$ along with the expression (\ref{SelfLuttinger}) for the derivative of the Luttinger-Ward functional, define the Legendre
transform of the Luttinger-Ward functional. It is easy to verify that%

\begin{equation}
\frac{\delta F[\Sigma]}{\delta\Sigma}=\frac{\delta\Phi[G]}{\delta
G}\frac{\delta G[\Sigma]}{\delta\Sigma}-\Sigma\frac{\delta G[\Sigma]}{\delta\Sigma}-G=-G
\end{equation}

hence, $\Omega_{\mathbf{t}}[\Sigma]$ is stationary with respect to $\Sigma$
when Dyson's equation is satisfied%
\begin{equation}
\frac{\delta\Omega_{\mathbf{t}}[\Sigma]}{\delta\Sigma}=-G+(G_{0\mathbf{t}}%
^{-1}-\Sigma)^{-1}=0.
\end{equation}

We now take advantage of the fact that $F[\Sigma]$  is universal, i.e., that it depends only on the interaction part of the Hamiltonian and not on the one-body part. This follows from the universal character of its Legendre transform $\Phi[G]$. We thus evaluate $F[\Sigma]$ exactly for a Hamiltonian $H^{\prime}$ that shares the same interaction part as the Hubbard Hamiltonian, but that is exactly solvable. This Hamiltonian $H^{\prime}$ is taken as a cluster decomposition of the original problem, i.e., we tile the infinite lattice into identical, disconnected clusters that can be solved exactly. Denoting the corresponding quantities with a prime, we obtain,
\begin{equation}
\Omega_{\mathbf{t}^{\prime}}[\Sigma^{\prime}]=F[\Sigma^{\prime}]-\mathrm{Tr}%
\ln(-G_{0\mathbf{t}^{\prime}}^{-1}+\Sigma^{\prime}),
\end{equation}
from which we can extract $F[\Sigma^{\prime}]$. It follows that
\begin{equation}
\Omega_{\mathbf{t}}[\Sigma^{\prime}]=\Omega_{\mathbf{t}^{\prime}}%
[\Sigma^{\prime}]+\mathrm{Tr}\ln(-G_{0\mathbf{t}^{\prime}}^{-1}+\Sigma
^{\prime})-\mathrm{Tr}\ln(-G_{0\mathbf{t}}^{-1}+\Sigma^{\prime}).
\label{sef_eq}%
\end{equation}
The fact that the self-energy (real and imaginary parts)
$\Sigma^{\prime}$ is restricted to the exact self-energy of the cluster
problem $H^{\prime}$, means that variational parameters appear in the definition
of the one-body part of $H^{\prime}$.

In practice, we look for values of the cluster one-body parameters
$\mathbf{t}^{\prime}$ such that $\delta\Omega_{\mathbf{t}}[\Sigma^{\prime
}]/\delta\mathbf{t}^{\prime}=0$. It is useful for what follows to write the
latter equation formally, although we do not use it in actual calculations.
Given that $\Omega_{\mathbf{t}^{\prime}}[\Sigma^{\prime}]$ is the grand
potential evaluated for the cluster, $\partial\Omega_{\mathbf{t}^{\prime}%
}[\Sigma^{\prime}]/\partial\mathbf{t}^{\prime}$ is cancelled by the explicit
$\mathbf{t}^{\prime}$ dependence of $\mathrm{Tr}\ln(-G_{0\mathbf{t}^{\prime}%
}^{-1}+\Sigma^{\prime})$ and we are left with
\begin{align}
0  &  =\frac{\delta\Omega_{\mathbf{t}}[\Sigma^{\prime}]}{\delta\Sigma^{\prime
}}\frac{\delta\Sigma^{\prime}}{\delta\mathbf{t}^{\prime}}\nonumber\\
&  =-\mathrm{Tr}\left[\left(  \frac{1}{G_{0\mathbf{t}^{\prime}}^{-1}%
-\Sigma^{\prime}}-\frac{1}{G_{0\mathbf{t}}^{-1}-\Sigma^{\prime}}\right)
\frac{\delta\Sigma^{\prime}}{\delta\mathbf{t}^{\prime}}\right]  .
\end{align}
Given that the clusters corresponding to $\mathbf{t}^{\prime}$ are
disconnected and that translation symmetry holds on the superlattice of
clusters, each of which contains $N_c$ sites,
the last equation may be written
\begin{align}
&  \sum_{\omega_{n}}\sum_{\mu\nu}\bigg[\frac{N}{N_{c}}\left(  \frac
{1}{G_{0\mathbf{t}^{\prime}}^{-1}-\Sigma^{\prime}(i\omega_{n})}\right)
_{\mu\nu}\nonumber\\
&  ~~ -\sum_{\tilde{\mathbf{k}}}\left(  \frac{1}{G_{0\mathbf{t}}%
^{-1}(\tilde{\mathbf{k}})-\Sigma^{\prime}(i\omega_{n})}\right)  _{\mu\nu
}\bigg]\frac{\delta\Sigma_{\nu\mu}^{\prime}(i\omega_{n})}{\delta
\mathbf{t}^{\prime}}=0. \label{EulerVCA}%
\end{align}

\subsubsection{Cellular Dynamical Mean-Field Theory}

The Cellular dynamical mean-field theory (CDMFT)\index{CDMFT}~\cite{Kotliar:2001} is obtained by including in
the cluster Hamiltonian $H^{\prime}$ a bath of uncorrelated electrons that
somehow must mimic the effect on the cluster of the rest of the lattice.
Explicitly, $H^{\prime}$ takes the form
\begin{align}
H^{\prime} &  =-\sum_{\mu,\nu,\sigma}t_{\mu\nu}^{\prime}c_{\mu\sigma}%
^{\dagger}c_{\nu\sigma}+U\sum_{\mu}n_{\mu\uparrow}n_{\mu\downarrow}\nonumber\\
&  ~~+\sum_{\mu,\alpha,\sigma}V_{\mu\alpha}(c_{\mu\sigma}^{\dagger}%
a_{\alpha\sigma}+\mathrm{H.c.})+\sum_{\alpha}\epsilon_{\alpha}a_{\alpha\sigma
}^{\dagger}a_{\alpha\sigma}%
\end{align}
where $a_{\alpha\sigma}$ annihilates an electron of spin $\sigma$ on a bath
orbital labelled $\alpha$. The bath is characterized by the energy of each
orbital ($\epsilon_{\alpha}$) and the bath-cluster hybridization matrix
$V_{\mu\alpha}$. The effect of the bath on the electron
Green function is encapsulated in the so-called hybridization function
\begin{equation}
\Gamma_{\mu\nu}(\omega)=\sum_{\alpha}{\frac{V_{\mu\alpha}V_{\nu\alpha}^{\ast}%
}{\omega-\epsilon_{\alpha}}}%
\end{equation}
which enters the Green function as
\begin{equation}
[G^{\prime-1}]_{\mu\nu}=\omega+\mu-t_{\mu\nu}^{\prime}-\Gamma_{\mu\nu
}(\omega)-\Sigma_{\mu\nu}(\omega).
\end{equation}

Moreover, the CDMFT does not look for a strict solution of the Euler equation
(\ref{EulerVCA}), but tries instead to set each of the terms between brackets to zero separately. Since the Euler
equation (\ref{EulerVCA}) can be seen as a scalar product, CDMFT requires
that the modulus of one of the vectors vanish to make the scalar product
vanish. From a heuristic point of view, it is as if each component of the
Green function in the cluster were equal to the corresponding component
deduced from the lattice Green function. This clearly reduces to single site
DMFT when there is only one lattice site.

\index{periodization}Clearly, in this approach we have lost translational invariance. The self-energy and Green functions depends not only on the superlattice wave vector $\tilde{\mathbf{k}}$, but also on cluster indices. By going to the Fourier space labeled by as many $\mathbf{K}$ values as cluster indices, the self-energy or the Green function may be written as functions of two momenta, for example $G(\tilde{\mathbf{k}}+\mathbf{K},\tilde{\mathbf{k}}+\mathbf{K'})$. In the jargon, periodizing a  function to recover translational-invariance, corresponds to keeping only the diagonal pieces, $\mathbf{K}=\mathbf{K'}$. The final lattice Green function from which one computes observable quantities is obtained by periodizing the self-energy,~\cite{Kotliar:2001} or the cumulants,~\cite{Stanescu:2006} or the Green function itself. The last approach can be justified within the self-energy functional mentioned above because it corresponds to the Green function needed to obtain the density from $\partial\Omega/\partial \mu=-\mathrm{Tr}(G)$. Periodization of the self-energy gives additional unphysical states in the Mott gap.~\cite{Senechal:2003}\footnote{There exists also a version of DMFT formulated in terms of cumulants Ref.~\cite{Stanescu:2004}} The fact that the cumulants are maximally local is often used to justify their periodization.~\cite{Stanescu:2006} Explicit comparisons of all three methods appear in Ref.~\cite{Sakai:2012}.

The DCA\cite{Hettler:1998}\index{DCA} cannot be formulated within the self-energy
functional approach.\footnote{Th. Maier, M. Potthoff and D. S\'{e}n\'{e}chal, unpublished.} It is based on the idea of discretizing irreducible quantities, such as the self-energy, in reciprocal space. It is believed to converge faster for $\mathbf{q=0}$ quantities whereas CDMFT converges exponentially fast for local quantities.\cite{Biroli:2002, Aryanpour:2005, Biroli:2005,KyungQMC:2006}

\subsubsection{Impurity solver}\index{impurity solver}

The problem of a cluster in a bath of non-interacting electrons is not trivial. It can be attacked by a variety of methods, ranging from exact diagonalization,~\cite{Caffarel:1994,Kancharla:2008,Capone:2004,CivelliThesis:2006,kyung:2006b,LiebschTong:2009,LiebschIshidaBathReview:2012} numerical renormalization group,~\cite{hofstetter_generalized_2000} to Quantum Monte Carlo~\cite{Hettler:1998}. The Continuous-Time Quantum-Monte-Carlo solver can handle an infinite bath and is the only one that is in principle exact, apart from controllable statistical uncertainties.~\cite{Gull:2011} 

For illustrative purposes, I briefly discuss the exact diagonalization solver introduced in Ref.~\cite{Caffarel:1994} in the
context of DMFT (i.e., a single site).~\footnote{For a pedagogical introduction, see~\cite{SenechalIntroductionQCM:2008,avella_cluster_2012}} When the bath is discretized, i.e., is made of a finite number of bath
\textquotedblleft orbitals\textquotedblright, the left-hand side of
Eq.~(\ref{EulerVCA}) cannot vanish separately for each frequency, since the
number of degrees of freedom in the bath is insufficient. Instead, one adopts
the following self-consistent scheme: (1) one starts with a guess value of the
bath parameters $(V_{\mu\alpha},\epsilon_{\alpha})$ and solves the cluster
Hamiltonian $H^{\prime}$ numerically. (2) One then calculates the combination
\begin{equation}
\hat{\mathcal{G}}_{0}^{-1}=\left[\sum_{\tilde{\mathbf{k}}}\frac
{1}{\hat{G}_{0\mathbf{t}}^{-1}(\tilde{\mathbf{k}})-\hat{\Sigma}^{\prime}(i\omega_{n}%
)}\right]  ^{-1}+\hat\Sigma^{\prime}(i\omega_{n})
\end{equation}
and (3) minimizes the following canonically invariant distance function
\begin{equation}
d=\sum_{n,\mu,\nu}\left\vert\left( i\omega_{n}+\mu-\hat{t}^\prime-\hat\Gamma
(i\omega_{n})-\hat{\mathcal{G}}_{0}^{-1}\right)_{\mu\nu}\right\vert ^{2}\label{dist_func}%
\end{equation}
over the set of bath parameters (changing the bath parameters at this step
does not require a new solution of the Hamiltonian $H^{\prime}$, but merely a
recalculation of the hybridization function $\hat{\Gamma}$). The bath
parameters obtained from this minimization are then put back into step (1) and
the procedure is iterated until convergence.

In practice, the distance function (\ref{dist_func}) can take various forms,
for instance by adding a frequency-dependent weight in order to emphasize
low-frequency properties\cite{Kancharla:2008,Bolech:2003,Stanescu:2005} or
by using a sharp frequency cutoff.~\cite{Kancharla:2008} These weighting factors
can be considered as rough approximations for the missing factor $\delta
\Sigma_{\nu\mu}^{\prime}(i\omega_{n})/\delta\mathbf{t}^{\prime}$ in the Euler
equation (\ref{EulerVCA}). The frequencies are summed over on a discrete,
regular grid along the imaginary axis, defined by some fictitious inverse
temperature $\beta$, typically of the order of 20 or 40 (in units of $t^{-1}%
$). 


    \subsection{Normal state and the pseudogap}
    Close to half-filling, as we discussed above, the normal state of high-temperature superconductors exhibits special properties. Up to optimal doping roughly, there is a doping dependent temperature, $T^*$ where a gap slowly opens up as temperature is decreased. This phenomenon is called a "pseudogap". We have discussed it briefly above. $T^*$ decreases monotonically with increasing doping. The signature of the pseudogap is seen in many physical properties. For example, the uniform magnetic spin susceptibility, measured by the Knight shift in nuclear magnetic resonance,~\cite{Alloul:1989} decreases strongly with temperature, by contrast with an ordinary metal where the spin susceptibility, also known as Pauli susceptibility, is temperature independent. Also, the single-particle density of states develops a dip between two energies on either side of the Fermi energy whose separation is almost temperature independent.~\cite{renner_pseudogap_1998} Angle-Resolved-Photoemission (ARPES) shows that states are pushed away from the Fermi energy in certain directions.~\cite{loeser_excitation_1996,ding_angle-resolved_1996} To end this non-exhaustive list, we mention that the c-axis resistivity increases with decreasing temperature while the optical conductivity develops a pseudogap~\cite{puchkov_pseudogap_1996}.~\footnote{For a short review, see Ref.~\cite{norman:2005}. An older review appears in Ref.~\cite{Timusk:1999}.}

	There are three broad classes of mechanisms for opening a pseudogap. a) Since phase transitions often open up gaps, the pseudogap could appear because of a first-order transition rounded by disorder. b) In two-dimensions, the Mermin-Wagner-Hohenberg theorem prohibits the breaking of a continuous symmetry. However, there is a regime with strong fluctuations that leads to the opening of a precursor of the true gap that will appear in the zero-temperature ordered state. We briefly explained this mechanism at the end of Sec.\ref{MethodsNormalState} and its application to electron-doped cuprates, that are less strongly coupled than the hole-doped ones.~\cite{Senechal:2004,Weber:2010} Further details are in Ref.~\cite{TremblayMancini:2011}. c) Mott physics by itself can lead to a pseudogap. This mechanism, different from the previous ones, as emphasized before,~\cite{Senechal:2004,Hankevych:2006} is considered in the present section. As discussed at the end of Section \ref{SuperconductorsQualitative} and in Fig.~\ref{Fig:XAS}, the hole-doped cuprates are doped Mott insulators, so this last possibility for a pseudogap needs to be investigated. 

\begin{figure}[t!]
 \centering
 \includegraphics[width=0.8\textwidth]{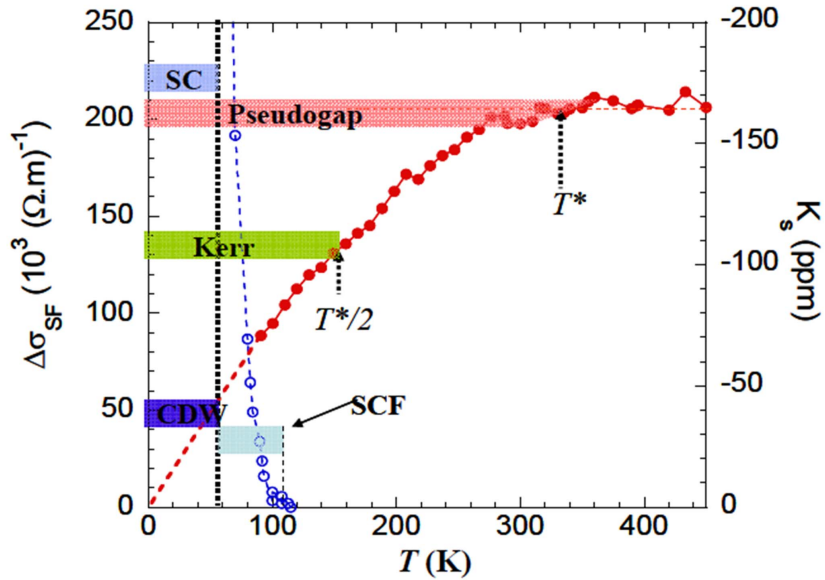}
 \caption{In red, the experimental spin contribution to the Knight shift (on the right vertical axis) for YBCO$_{6.6}$.~\cite{Alloul:1991} The illustrated phase transitions for superconducting transition temperature, for the Kerr signal~\cite{Kapitulnik:2008} and for the high-field CDW from NQR quadrupole effects~\cite{WuJulien:2011} and ultrasound velocity data~\cite{leboeuf_thermodynamic_2013}. The CDW transition from X-Ray diffraction in zero field~\cite{Chang:2012} coincides with the onset of the Kerr effect. Recent resonant ultrasound measurements, suggest a true phase transition~\cite{shekhter_bounding_2013} at $T^*$, like the Nernst signal~\cite{Daou:2010}. The sharp drop in the superconducting fluctuation conductivity (SCF)~\cite{AlloulEPL:2010} is also illustrated in blue with the corresponding vertical axis on the left. Figure from Ref.~\cite{Alloul:2013}.}
\label{Fig:Alloul}
\end{figure}
	
Before proceeding further, note that the candidates for the order parameter of a phase transition associated with the pseudogap are numerous: Stripes,~\cite{Kivelson:2003}, nematic order,~\cite{Daou:2010} d-density wave,~\cite{Chakravarty:2001} antiferromagnetism~\cite{ChangCompetingAFM:2008}... There is strong evidence in several cuprates of a charge-density wave~\cite{WuJulien:2011,Ghiringhelli:2013,Chang:2012} (anticipated from transport~\cite{TailleferCDW:2010} and quantum oscillations~\cite{Doiron:2007}) and of intra-unit cell nematic order~\cite{Lawler:2010}. All of this is accompanied by Fermi surface reconstruction.~\cite{TailleferReconstruction:2012,Doiron:2007,Sebastian:2012} Time-reversal symmetry breaking also occurs, as evidenced by the Kerr effect~\cite{Kapitulnik:2008} and by the existence of intra-unit cell spontaneous currents as evidenced by polarized neutron scattering.~\cite{SidisReviewTR:2013,Varma:1997} Nevertheless, it seems that these order appear at lower temperature than the pseudogap temperature.~\cite{Alloul:2013} The orders seem a consequence rather than the cause of the pseudogap.~\cite{Gomes:2007} This is illustrated by Fig.~\ref{Fig:Alloul} and discussed further in the corresponding caption. Clearly, some of these orders, such as intra-unit cell spontaneous currents,~\cite{Varma:1997,SidisReviewTR:2013} cannot be explained within a one-band model. Nevertheless, since the order generally appears below the $T^*$ illustrated in the figure, (see however caption of Fig.~\ref{Fig:Alloul}) it is worth investigating the predictions of the simple one-band model.

Early finite temperature DCA~\cite{Huscroft:2001}, and zero-temperature exact diagonalizations with Cluster Perturbation Theory~\cite{Senechal:2004} and  with CDMFT~\cite{kyung:2006b,Haule:2007,Kancharla:2008,LiebschTong:2009,Ferrero:2009}, have shown that the calculated pseudogap for ARPES is very close to experiment. Several recent calculations with CDMFT~\cite{Haule:2007,GullFerrero:2010,Sordi:2012} or DCA~\cite{Werner:2009,GullFerrero:2010} using Continuous-Time Quantum Monte Carlo at finite temperature as an impurity solver have found similar results. 

\begin{figure}[t!]
 \centering
 \includegraphics[width=0.49\textwidth]{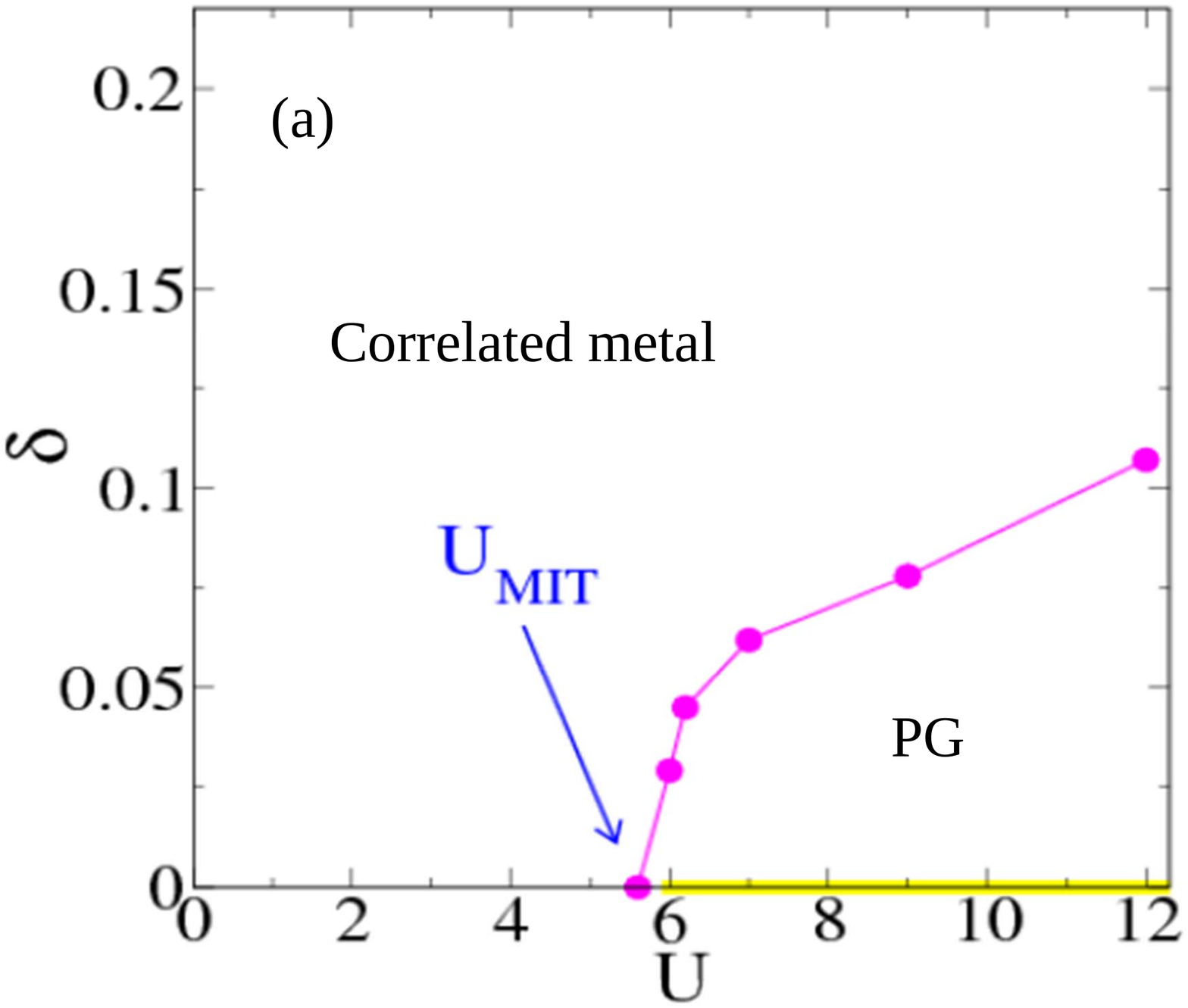}
 \includegraphics[width=0.49\textwidth]{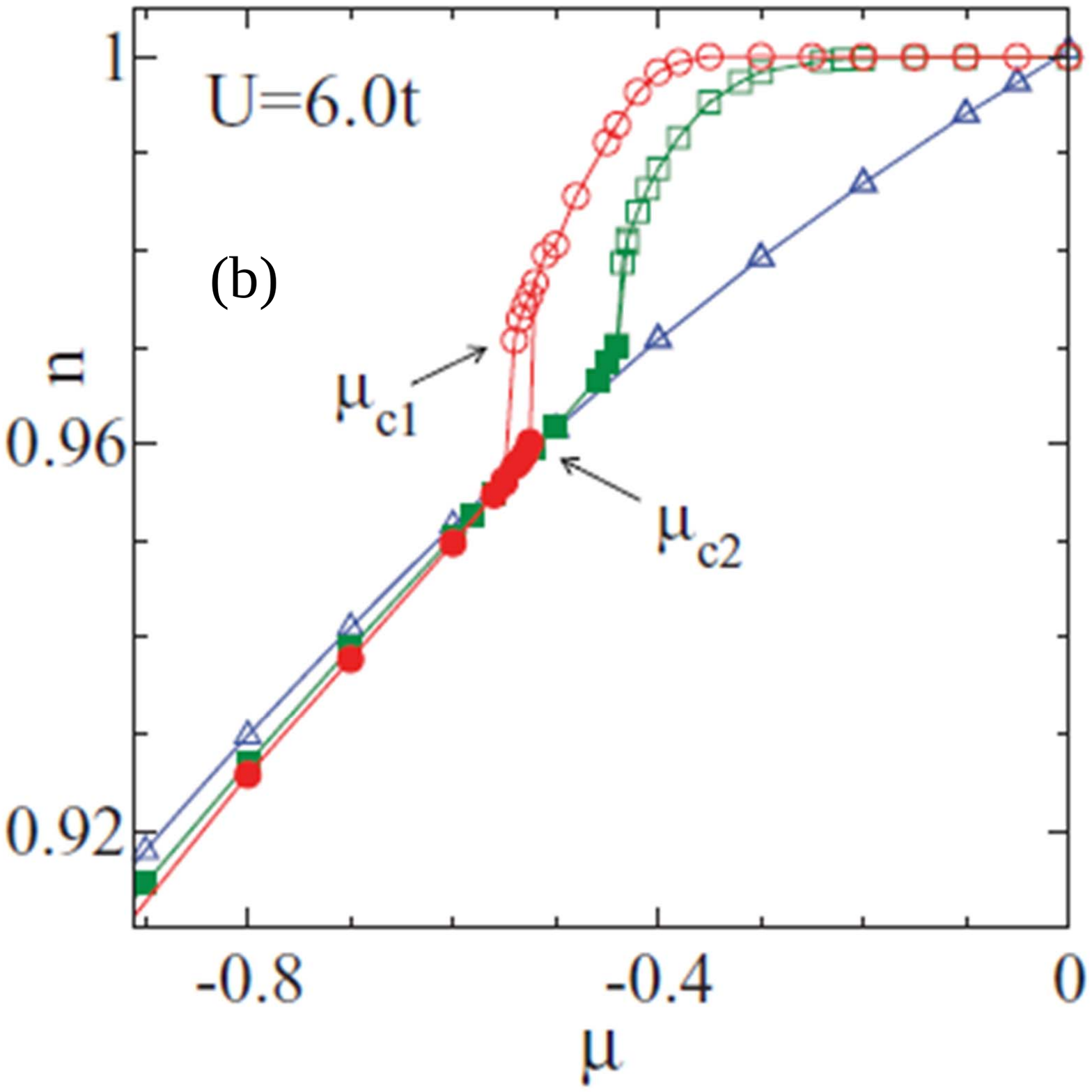}
 \caption{(a) Zero-temperature extrapolation of the normal-state phase diagram for a $2\times 2$ plaquette for the Hubbard model with nearest-neighbor hopping. In addition to the Mott transition at zero doping where the insulating phase is in yellow, there is a first-order transition line that separates a phase with a pseudogap from a strongly correlated metal. Figure adapted by G. Sordi from ~\cite{Sordi:2011}. Signs of a first order transition have also been seen in Ref.~\cite{LiebschTong:2009} (b) Filling $n$ versus chemical potential $\mu$ for different temperatures: At high temperature, $T=1/10$, there is a single curve represented by the blue line with triangles. At lower temperatures, $T=1/25$ in green with squares, and at $T=1/50$ in red with circles, there are clear signs of a first order transition.~\cite{Sordi:2010,Sordi:2011}}
\label{Fig:PhaseDiagram}
\end{figure}

Fig.~\ref{Fig:PhaseDiagram} illustrates the essential features of the normal state phase diagram for the Hubbard model with nearest-neighbor hopping only on a $2\times 2$ plaquette.~\cite{Sordi:2010,Sordi:2011} When the doping $\delta$ vanishes in Fig.~\ref{Fig:PhaseDiagram}(a), the insulating phase, represented by the yellow line, begins around $U=5.8$. Hysteresis (not shown) occurs as a function of $U$. From this line, emerges another first-order transition line that separates two types of metals: A metallic state with a pseudogap near half-filling, and a correlated metal away from it. The transition between the two metals is well illustrated in Fig.~\ref{Fig:PhaseDiagram}(b). Consider the $n(\mu)$ curve at $T=1/50$ represented by circles on a red line. Decreasing $\mu$ from $\mu=0$, the density remains fixed at $n=1$ for a relatively large range of $\mu$ because the Mott gap is opened and the chemical potential is in the gap. Around $\mu=-0.4$ one enters a compressible phase, i.e. $dn/d\mu$ finite. The rounded crossover is due to the finite temperature. It should become a discontinuous change in slope at $T=0$. The jump in filling and the hysteresis is obvious near $\mu=-0.55$. 

\begin{figure}[t!]
 \centering
 \includegraphics[width=1.0\textwidth]{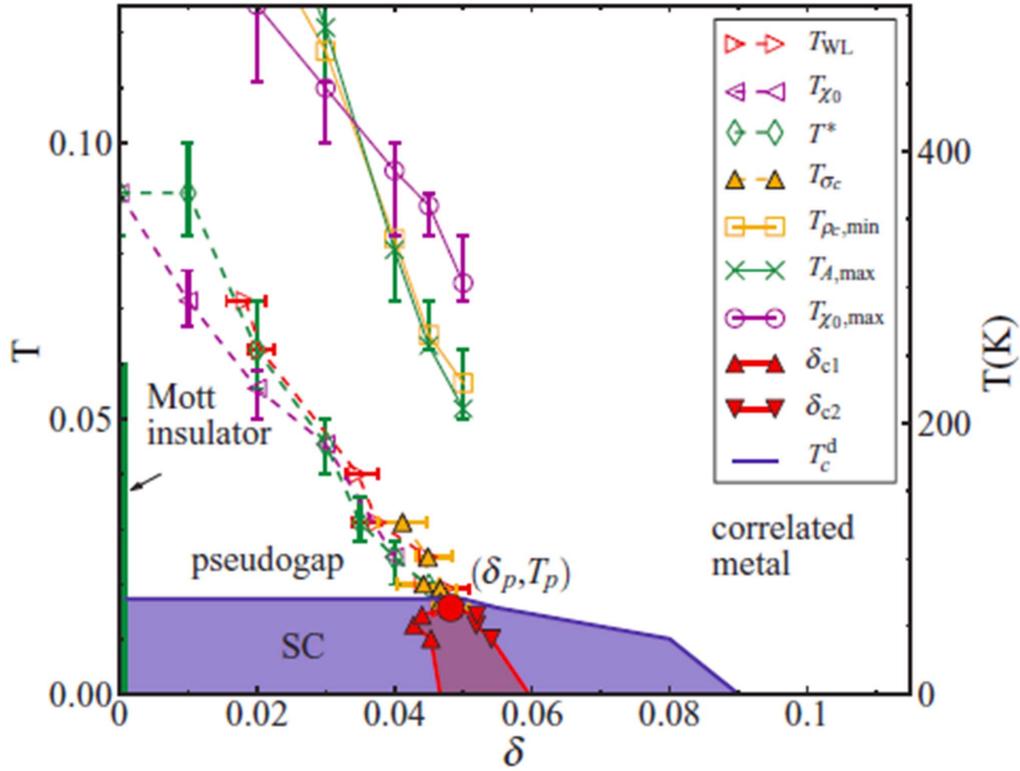}
 \caption{Temperature vs doping phase diagram for the nearest-neighbor Hubbard model for $U=6.2$. The vertical axis on the left-hand side is in units of hopping. The light blue region delineates the superconducting phase. In the absence of that phase, i.e. in the metastable normal state, one finds a first-order coexistence region between the two red lines terminating at the critical point $(\delta_p,T_p)$. The pseudogap phase is near half-filling. Red right-pointing arrows, maximum of the charge compressibility (Widom line); purple left-pointing arrows, inflection point of the spin susceptibility; green diamonds ($T^*$), inflection point in the zero-frequency local density of states; orange triangles: inflection point in $\sigma_c(\mu)$; orange squares, minimum in the c-axis resistivity; green crosses, maximum in the zero frequency density of states; purple circles, maximum of the spin susceptibility. Figure taken from Ref.~\cite{SordiResistivity:2013}}
\label{Fig:Widom}
\end{figure}

Let us now fix the interaction strength to $U=6.2$ and look at the phase diagram in the doping-temperature plane in Fig.~\ref{Fig:Widom}. This is obtained for the normal state, without allowing for antiferromagnetism (Disregard for the moment the superconducting region in blue. It will be discussed in the next subsection \footnote{Recent progress on the ergodicity of the hybridization expansion for continuous-time quantum Monte Carlo leads to results for the superconducting phase that are qualitatively similar but quantitatively different from those that appear for this phase in Figs.~\ref{Fig:Widom} and \ref{Fig:Misc}(c). These results will appear later.~\cite{SordiSemon:2013}}).~\cite{SordiSuperconductivityPseudogap:2012} The first-order transition is illustrated by the shaded region between the red lines. Various crossovers are identified as described in the caption. Let us focus on the two crossover lines associated with the uniform magnetic susceptibility. The purple solid line with circles that appears towards the top of the graph identifies, for a given doping, the temperature at which the susceptibility starts to fall just after it reaches a broad maximum. This was identified as $T^*$ in Fig.~\ref{Fig:Alloul}.\footnote{Note that $T^*$ in Fig.~\ref{Fig:Widom} refers to the inflection point of the zero frequency density of states as a function of temperature. It differs from $T^*$ in Fig.~\ref{Fig:Alloul}.} The purple line ends before it reaches zero temperature because at larger doping the maximum simply disappears. One recovers a Pauli-like susceptibility at these dopings.\footnote{The lines at high temperature end before zero temperature. They should not be extrapolated to zero temperature as is often done in other theoretical work.} The purple dashed line with left-pointing arrows identifies, at lower temperature, the inflection point of the susceptibility. It is very close to other crossover lines that all originate at the critical point $(\delta_p,T_p)$ of the first-order transition. The appearance of many crossover lines that all merge to terminate at the critical point is a very general phenomenon in phase transitions. It has been recently called a Widom line in the context of supercritical fluids~\cite{XuStanleyWidom:2005} and G. Sordi has suggested that the same concept applies for the electron fluid.~\cite{Sordi:2012}  In the latter case, a thermodynamic quantity, namely the compressibility, has a maximum at the right-pointing arrows along the red dashed line and dynamical quantities, such as the density of states, the conductivity and so on, have rapid crossovers, all analogous to the supercritical fluid case.

It is noteworthy that above the crossover line where a c-axis resistivity minimum occurs (orange squares in Fig.~\ref{Fig:Widom}) the temperature dependence is almost linear. In that regime, the c-axis resistivity can exceed the appropriate version of the Mott-Ioffe-Regel criterion.~\cite{ItoRhoC:1991} Also, at zero-temperature in the pseudogap phase, it was demonstrated that very small orthorhombicity leads to very large conductivity anisotropy in a doped Mott insulator. This called electronic dynamical nematicity.~\cite{Okamoto:2010} Such very large anisotropy is observed experimentally in YBCO.~\cite{Ando:2002} 

From the microcopic point of view, the probability that the plaquette is in a singlet state with four electrons increases rapidly as temperature decreases, reaching values larger than $0.5$ at the lowest temperatures. The inflection point as a function of temperature of the probability for the plaquette singlet coincides with the Widom line~\cite{Sordi:2012}. 

From the point of view of this analysis, the pseudogap is a distinct phase, separated from the correlated metal by a first-order transition. It is an unstable phase at low temperature since it appears only if we suppress antiferromagnetism and superconductivity. Nevertheless, as in the case of the Mott transition, the crossovers at high-temperatures are remnants of the first-order transition. The phase transition is in the same universality class as the liquid-gas transition. As $U$ increases, the critical point moves to larger doping and lower temperature. These calculations were for values of $U$ very close to the Mott transition so that one could reach temperatures low enough that the first-order transition is visible. Although one sees crossover phenomena up to very large values of $U$, the possibility that the first-order transition turns into a quantum critical point cannot be rejected.

	In summary for this section, one can infer from the plaquette studies that even with antiferromagnetic correlations that can extend at most to first-neighbor, one can find a pseudogap and, as I discuss below, d-wave superconductivity. The pseudogap mechanism in this case is clearly related to short-range Mott physics, 
not to AFM correlation lengths that exceed the thermal de Broglie wavelength. Similarly, the pairing comes from the exchange interaction $J$, but that does not necessarily mean long wavelength antiferromagnetic correlations. 

%

    \subsection{Superconducting state}


\begin{figure}[t!]
 \centering
 \includegraphics[width=1.0\textwidth]{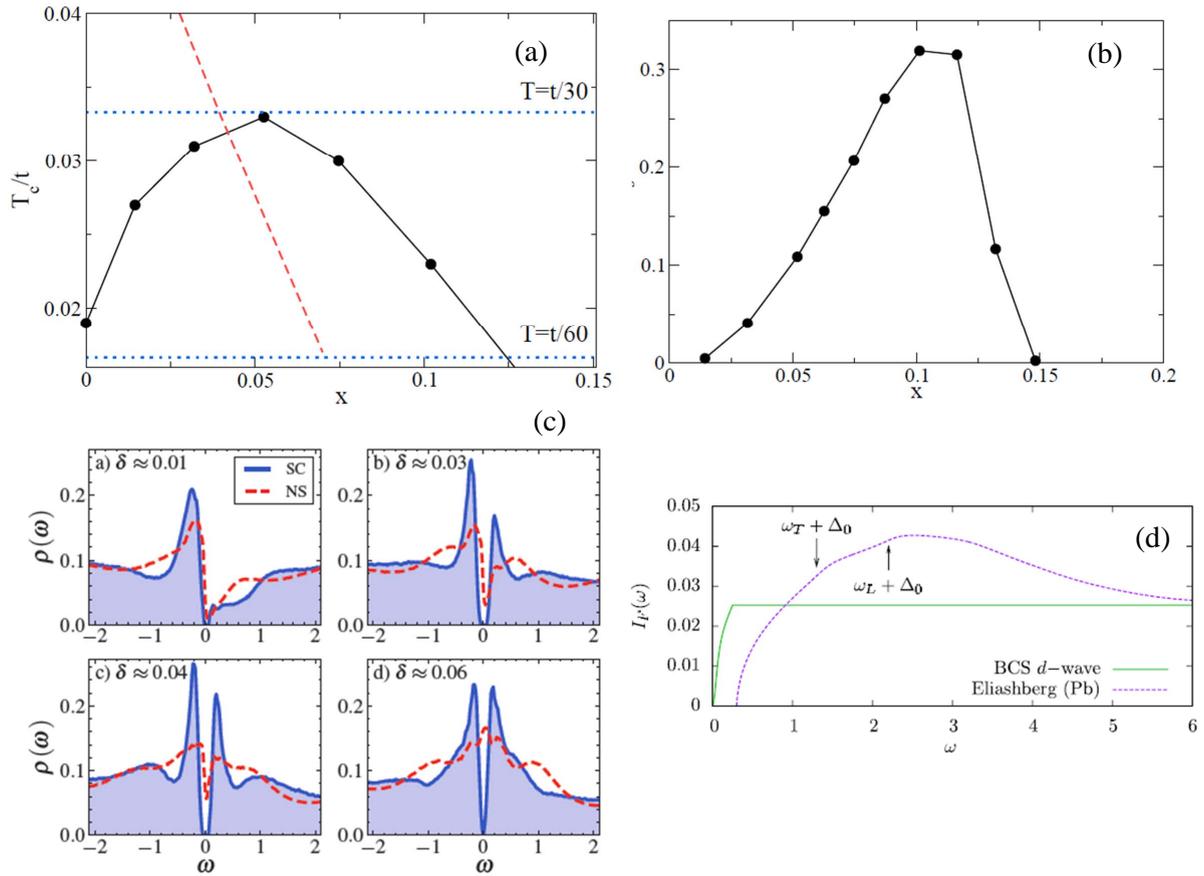}
 \caption{(a) Superconducting critical temperature of the Hubbard model with nearest neighbor hopping calculated for $U = 6t$ using the 8-site DCA.~\cite{gull_superconductivity_2013} Dashed red line denotes a crossover to the normal state pseudogap. Dotted blue lines indicate the range of temperatures studied. Note that the temperature axis does not begin at zero. Figure from Ref.~\cite{gull_superconductivity_2013} (b) Superfluid stiffness at $T = t/60$. Figure from Ref.~\cite{gull_superconductivity_2013} (c) Low frequency part of the local density of states $\rho (\omega)$ at $U=6.2t$, $T=1/100$ for the normal state and the superconducting state (red dashed and blue solid lines). Figure from Ref.~\cite{SordiSuperconductivityPseudogap:2012} (d)  Cumulative order parameter, i.e. the integral of the anomalous Green's function (or Gork'ov function) $I_F(\omega)$. The dashed green line is $I_{F}\left(\omega\right)$ for a d-wave BCS superconductor with a cutoff at $\omega_{c}=0.5$, which plays the role of the Debye frequency. In that case, the indices $i,j$ in Eq.~\eqref{F(w)} are near-neighbor. The magenta line is extracted~\cite{Kyung:2009} from Eliashberg theory for Pb in Ref.~\cite{Scalapino:1966}. Frequencies in that case are measured in units of the transverse phonon frequency, $\omega_T$. The scale of the vertical axis is arbitrary. For that $s-wave$ superconductor, one takes $i=j$ in Eq.~\eqref{F(w)}. Figure from Ref.~\cite{Kyung:2009}.}
\label{Fig:Misc}
\end{figure}  
	
	When the interaction $U$ is larger than that necessary to lead to a Mott insulator at half-filling, d-wave superconductivity has many features that are very non-BCS like. That is the topic of this section. 

	But first, is there d-wave superconductivity in the one-band Hubbard model or its strong-coupling version, the $t-J$ model? Many-methods suggest that there is~\cite{Paramekanti:2004,SorellaSC:2002,Poilblanc:2002,Varma:2013,monthoux_superconductivity_2007}. But there is no unanimity.~\cite{Aimi:2007} For reviews, see for example Refs.~\cite{maiti_superconductivity_2013,ScalapinoThread:2010,LTP:2006}. In DCA with large clusters and finite-size study, Jarrell's group has found convincing evidence of d-wave superconductivity~\cite{maier_d:2005} at $U=4t$. This is too small to lead to a Mott insulator at half-filling but even for $U$ below the Mott transition there was still sometimes some dispute regarding the existence of superconductivity. For larger $U$ and 8-site clusters~\cite{gull_superconductivity_2013} one finds d-wave superconductivity and pseudogap. It is remarkable that very similar results were obtained earlier with the variational cluster approximation on various size clusters~\cite{Senechal:2005,AichhornAFSC:2006,Aichhorn:2007} and with CDMFT on $2\times 2$ plaquettes.~\cite{Kancharla:2008,Haule:2007,SordiSuperconductivityPseudogap:2012} With a realistic band-structure, the competition between superconductivity and antiferromagnetism can be studied and the asymmetry between the hole and electron-doped cuprates comes out clearly.~\cite{Senechal:2005,AichhornAFSC:2006,Aichhorn:2007,Kancharla:2008}    

	Let us move back to finite temperature results. Fig.~\ref{Fig:Widom} shows that the superconducting phase appears in a region that is not delimited by a dome, as in experiment, but that instead the transition temperature $T_c^d$ becomes independent of doping in the pseudogap region.~\cite{SordiSuperconductivityPseudogap:2012} There is a first-order transition to the Mott insulator at half-filling. One expects that a mean-field treatment overestimates the value of $T_c^d$. That transition temperature $T_c^d$ in the pseudogap region is interpreted as the temperature at which local pairs form. This could be the temperature at which a superconducting gap appears in tunneling experiments~\cite{Gomes:2007,Gomes2008} without long-range phase coherence. Many other experiments suggest the formation of local pairs at high temperature in the pseudogap region.~\cite{PushpYazdaniConstantNodes:2009,DurokaKeimerLocalPairs:2011,kondo_disentangling_2011,rourke_phase-fluctuating_2011,Morenzoni_meissner_2011} Note however that $T_c^d$ is not the same as the pseudogap temperature. {\it The two phenomena are distinct~\cite{SordiSuperconductivityPseudogap:2012}}. 

	It is important to realize the following non-BCS feature of strongly-correlated superconductivity. The saturation of $T_c^d$ at low temperature occurs despite the fact that the order parameter has a dome shape, vanishing as we approach half-filling.~\cite{Kancharla:2008,SordiSuperconductivityPseudogap:2012} The order parameter is discussed further below. For now, we can ask what is the effect of the size of the cluster on $T_c^d$. Fig.~\ref{Fig:Misc}(a) for an 8 site cluster~\cite{gull_superconductivity_2013} shows that $T_c^d$ at half-filling is roughly $30\%$ smaller than at optimal doping, despite the fact that the low temperature superfluid density vanishes at half-filling, as seen in Fig.~\ref{Fig:Misc}(b). Again, this is not expected from BCS. Since it seems that extremely large clusters would be necessary to observe a dome shape with vanishing $T_c^d$ at infinitesimal doping, it would be natural to conclude that long wavelength superconducting or antiferromagnetic fluctuations are necessary to reproduce the experiment. The long-wavelength fluctuations that could be the cause of the decrease of $T_c^d$ could be quantum and classical phase fluctuations~\cite{EmeryKivelsonBad:1995,emery_importance_1995,PodolskyNerstPhaseFluctuations:2007,Tesanovic_d-wave_2008}
fluctuations in the magnitude of the order parameter~\cite{Ussishkin:2002}
or of some competing order, such as antiferromagnetism or charge-density wave. 
Evidently, the establishment of long-range order of a competing phase would also be effective.~\cite{Fradkin:2010}
Finally, in the real system, disorder can play a role~\cite{AlbenqueAlloul:2008,AlloulRMP:2009}

Another non-BCS feature of strongly correlated superconductivity appears in the single-particle density of states.~\cite{SordiSuperconductivityPseudogap:2012} Whereas in BCS the density of states is symmetrical near zero frequency, Fig.~\ref{Fig:Misc}(c) demonstrates that the strong asymmetry present in the pseudogap normal state (dashed red line) survives in the superconducting state. The asymmetry is clearly a property of the Mott insulator since it is easier to remove an electron ($\omega<0$) than to add one ($\omega>0$). Very near $\omega=0$, all the densities of states in Fig.~\ref{Fig:Misc}(c) are qualitatively similar since they are dictated by the symmetry-imposed nodes nodes that are an emergent property of d-wave superconductors. Once the normal state is a correlated metal, for example at doping $\delta=0.06$, the (particle-hole) symmetry is recovered. 

The correlated metal leads then to superconducting properties akin to those of spin-fluctuation mediated BCS superconductivity. For example, in the overdoped regime superconductivity disappears concomitantly with the low frequency peak of the local spin susceptibility.~\cite{Kyung:2009} But in the underdoped regime where there is a pseudogap, the difference between the pairing mechanism in a doped Mott insulator and the pairing mechanism in a doped fluctuating itinerant antiferromagnet comes out very clearly when one takes into account nearest-neighbor repulsion.~\cite{SenechalResilience:2013} Indeed, the doped Mott insulator is much more resilient to near-neighbor repulsion than a spin-fluctuation mediated BCS superconductor, for reasons that go deep into the nature of superconductivity in a doped Mott insulator. This is an important result that goes much beyond the mean-field arguments of Eqs.~\eqref{t-J-Model} to \eqref{H_MF}. In this approach, when there is near-neighbor repulsion, one finds that superconductivity should disappear when $V>J$. In cuprates, taking the value of the near-neighbor Coulomb interaction with a relative dielectric constant of order 10 we estimate that $V$, the value of near-neighbor repulsion, is of order $V \approx 400$~meV while $J \approx 130$~meV. So, from the mean-field point of view, superconductivity would not occur in the hole-doped cuprates under such circumstances. 


	\index{cumulative order parameter} To understand the resilience of strongly correlated superconductivity to near-neighbor repulsion $V$, we need worry about the dynamics of pairing. To this end, consider
the function $I_{F}(\omega)$ defined through the integral
\begin{equation}
I_{F}(\omega) = -\int_0^\omega\frac{\mathrm{d}\omega'}{\pi}
\operatorname{Im}F_{ij}^{R}(\omega')
\label{F(w)}
\end{equation}
where $F^{R}$ is the retarded Gork'ov function (or off-diagonal Green's function in the Nambu formalism) defined in imaginary time by $F_{ij}\equiv-\langle Tc_{i\uparrow}(\tau)c_{j\downarrow}(0)\rangle$ with $i$ and $j$ nearest-neighbors. The infinite frequency limit of $I_{F}(\omega) $ is equal to $\left\langle c_{i\uparrow}c_{j\downarrow}\right\rangle $ which in turn is proportional to the $d$-wave order parameter $\psi$. As should become clear below, $I_{F}(\omega)$ is useful to estimate the frequencies that are relevant for pair binding. The name ``cumulative order parameter'' for $I_{F}(\omega)$~\cite{SenechalResilience:2013} is suggestive of the physical content of that function.

	Fig.~\ref{Fig:Misc}(d) illustrates the behavior of $I_{F}\left(\omega\right)$ in well known cases. The dashed green line is $I_{F}\left(\omega\right)$ for a d-wave BCS superconductor with a cutoff at $\omega_{c}=0.5.$ In BCS theory, that would be the Debye frequency. In BCS then, $I_F(\omega)$ is a monotonically increasing function of $\omega$ that reaches its asymptotic value at the BCS cutoff frequency $\omega_c$.~\cite{Kyung:2009}. The magenta line in Fig.~\ref{Fig:Misc}(d) is obtained from Eliashberg theory for Pb in Ref.~\cite{Scalapino:1966}. The two glitches before the maximum correspond to the transverse, $\omega_T$, and longitudinal, $\omega_L$, peaks in the phonon density of states. In the Eliashberg approach that includes a retarded phonon interaction as well as the Coulomb pseudopotential $\mu^*$ that represents the repulsive electron-electron interaction,~\cite{Kyung:2009} the function overshoots its asymptotic value at frequencies near the main phonon frequencies before decaying to its final value. The decrease occurs essentially because of the presence of the repulsive Coulomb pseudopotantial, as one can deduce~\cite{SenechalResilience:2013} from the examples treated in Ref.~\cite{Scalapino:1966}.
	 
\begin{figure}[t!]
 \centering
 \includegraphics[width=1.0\textwidth]{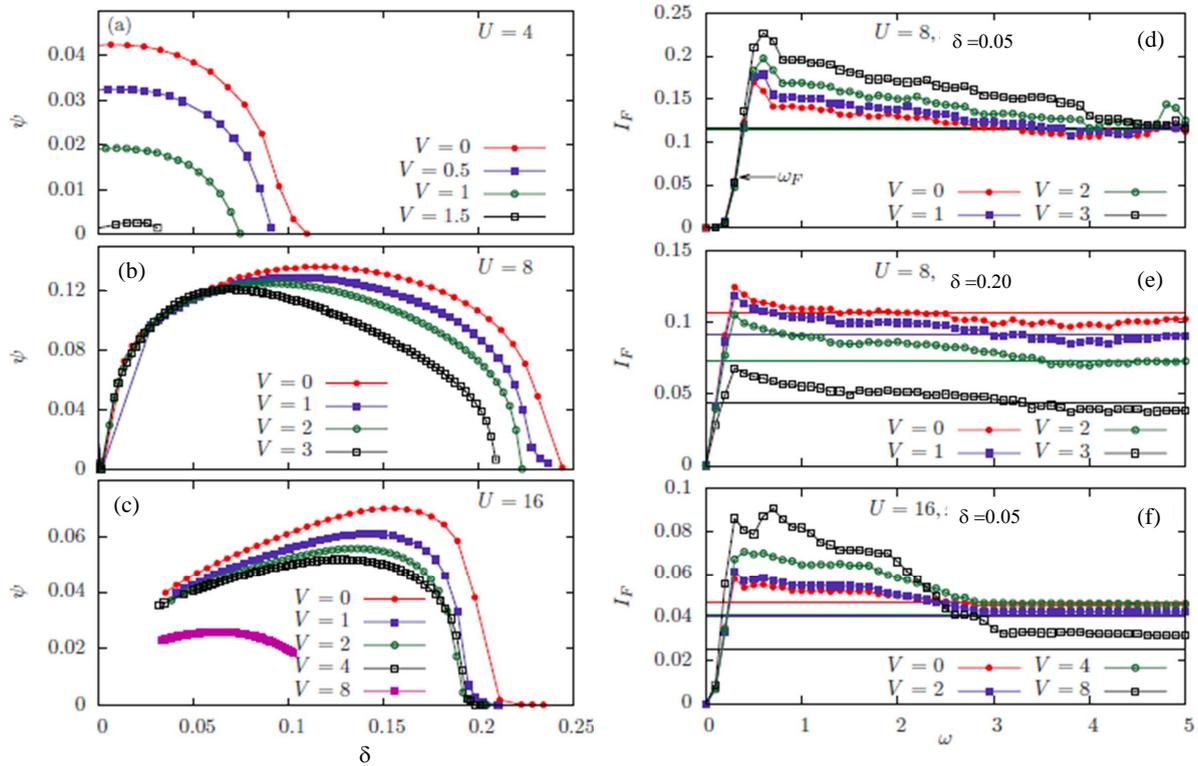}
 \caption{All results for these figures were obtained with CDMFT and the exact diagonalization solver with the bath parametrization defined in Ref.~\cite{SenechalResilience:2013}. The three panels on the left are for the $d$-wave order parameter $\psi$ obtained from the off-diagonal component of the lattice Green function as a function of cluster doping for (a) $U=4t$, (b) $U=8t$ and (c) $U=16t$ and various values of $V$. The three panels on the right represent the integral of the anomalous Green function (or Gork'ov function) $I_F(\omega)$ obtained after extrapolation to $\eta=0$ of $\omega+i\eta$ for several values of $V$ at (d) $U=8t$, $\delta=0.05$ (e) $U=8t$, $\delta=0.2$ (f) $U=16t$, $\delta=0.05$ with $\delta$ the value of doping. Frequency is measured in energy units with $t=1$. The asymptotic value of the integral, $I_F(\infty)$, equal to the order parameter, is shown as horizontal lines. $I_F(\omega)$ is the cumulative order parameter defined by Eq.~\eqref{F(w)}. The characteristic frequency $\omega_F$ is defined as the frequency at which $I_F(\omega)$ is equal to half of its asymptotic value. The horizontal arrow in panel (d) indicates how $\omega_F$ is obtained.}
\label{Fig:d-waveOP}
\end{figure}

	The resilience of strongly correlated superconductors to near-neighbor repulsion is best illustrated by Figs.~\ref{Fig:d-waveOP}(a) to (c).	Each panel illustrates the order parameter as a function of doping. The dome shape that we alluded to earlier appears in panels (b) and (c) for $U$ larger than the critical value necessary to obtain a Mott insulator at half-filling. At weak coupling, $U=4t$ in panel (a), the dome shape appears if we allow antiferromagnetism to occur.~\cite{Kancharla:2008} Panel (a) illustrates the sensitivity of superconductivity to near-neighbor repulsion $V$ at weak coupling. At $V/U=1.5/4$ superconductivity has disappeared. By contrast, at strong coupling, in panel (b) $U=8t$, one notices that for $V$ twice as large as in the previous case and for the same ratio $V/U=3/8$ superconductivity is still very strong. In fact, the order parameter is not very sensitive to $V$ in the underdoped regime. Sensitivity to $V$ occurs mostly in the overdoped regime, which is more BCS-like. The same phenomena are observed in panel (c) for $U=16t$.  

	To understand the behavior of the order parameter in the strongly-correlated case, let us return to the cumulative order parameter. If we define the characteristic frequency $\omega_F$ as the frequency at which the cumulative order parameter reaches half of its asymptotic value, one can check that this frequency scales as $J$. Hence, not only does the optimal value of the order parameter scale as $J$, but so does the characteristic frequency over which the order parameter builds up. 
	
	The qualitative aspects of the effects of $V$, are clearer if we focus on the largest value $U=16t$ in panel (f) for the underdoped case $\delta=0.05$. The red curve with filled circles is for $V=0$. For $V=8t$, the black curve with open squares shows that the maximum of the cumulative order parameter is larger by roughly a factor of 2 than the $V=0$ case. This is because at strong coupling, $V$ also contributes to the effective $J$ and thus to attraction at low frequencies. Indeed, in the presence of $V$, $J$ increases from $4t^2/U$ at $V=0$ to $4t^2/(U-V)$. That increase can be understood by considering two singly-occupied neighboring spins on a square lattice. If all other sites are occupied, the contribution to the potential energy from $V$ is $7V$. When one of the two electrons is placed on the same site as its neighbor, they together now have three neighbors, hence the potential energy contribution from $V$ is $6V$. The energy denominator in second-order degenerate perturbation theory thus becomes $U-7V+6V=U-V$, which explains the increased value of $J$ in the presence of $V$. The ratio of the effective $J$ at $U=16t$, $V=8t$ to that at $U=16t$, $V=0$ is $U/(U-V)=2$, which explains the observed increase in the maximum value of $I_F(\omega)$ by a factor of two. At larger frequencies however, $V$ is detrimental to superconductivity, leading to a Coulomb pseudopotential $\mu^*$ that reduces the value of the order parameter, just as in the case of lead described above. The effect of $\mu^*$ is largest when $V$ is largest, as expected. Overall, the low-frequency contribution of $J$ to the increase in binding is compensated by the high-frequency detrimental effect of $V$. 
	
	It is also remarkable that the frequency scale over which the cumulative order parameter reaches its asymptotic value seems to equal a few times $J$, just as it equals a few times the largest phonon frequency in Fig.~\ref{Fig:Misc}(d) for the Eliashberg theory of lead. This observation is consistent with the fact that the order parameter reaches its asymptotic value for $U=16t$ in Fig~\ref{Fig:d-waveOP}(f) at a frequency roughly half that where the asymptotic regime is reached for $U=8t$ in Fig~\ref{Fig:d-waveOP}(d). Indeed, at $V=0$, $J$ is smaller by a factor of two when $U$ increases by a factor of two.
	
	 By comparing Fig~\ref{Fig:Misc}(d) for the BCS and Eliashberg cases with Fig.~\ref{Fig:d-waveOP}(e) for the overdoped case $\delta=0.20$ and Fig.~\ref{Fig:d-waveOP}(c) for the underdoped case $\delta=0.05$, both for $U=8t$, one verifies that the overdoped case is more BCS-like. This is consistent with the greater sensitivity of the order parameter to $V$ that one can observe in the overdoped regime of Figs.~\ref{Fig:d-waveOP}(b) and (c).


	Let us end with one of the most striking properties of strongly-correlated superconductivity. Layered organic superconductors of the $\kappa$-BEDT family can be modelled by the one-band Hubbard model on an anisotropic triangular lattice at half-filling.~\cite{KinoFukuyama:1996,Powell:2006} By changing pressure, one can tune the normal state through a Mott transition. The metallic state is at high pressure. At low temperature, pressure changes the insulating state, that can be either antiferromagnetic~\cite{Lefebvre:2000} or spin liquid,~\cite{Shimizu:2003} to a superconducting state with a non s-wave order parameter. (Ref.~\cite{PowellMcKenzieReview:2011} for a review) One finds experimentally that the superconducting $T_c$ is largest at the transition, in other words, closest to the insulating phase. In addition, $T_c$ is larger in the compounds that have the largest mass renormalization in the normal metallic state, in other words in the most strongly correlated ones.~\cite{Powell:2006} All of this is highly counter-intuitive. I have run out of space to explain the theoretical situation on this issue. Let us just remark that at $T=0$ with an exact diagonalization solver, one finds with quantum cluster methods~\cite{Sahebsara:2006} that indeed the order-parameter is largest near the first-order Mott transition.~\cite{Kyung2006} If $T_c$ scales with the order parameter, this type of approach thus reproduces a counter-intuitive result. The existence of unconventional superconductivity in the Hubbard model on the anisotropic triangular lattice is however disputed.~\cite{ClayMazumdarNoSC:2012} 
	
	In the model for the cuprates, the value of the order parameter at optimal doping increases with $U$ starting from small $U$ and reaches a maximum at intermediate coupling before decreasing with $J$ at large $U$. In other words, as a function of $U$, the largest value that the order parameter can take is for $U\approx 6t$ where the Mott transition occurs at half-filling. Analogously, as a function of doping, at $U=6.2$, the maximum $T_c$ occurs near the critical doping for the pseudogap to correlated metal transition.~\cite{SordiSuperconductivityPseudogap:2012}  



\section{Conclusion}

Progress with numerical methods, especially cluster generalizations of DMFT, have shown in recent years that much of the physics of strongly correlated superconductors is contained in the one-band Hubbard model. Confidence in the method comes from extensive benchmarking, and from the agreement at intermediate coupling with TPSC, which is also benchmarked and valid up to intermediate coupling. The fact that both the cuprates and the layered organics are well described give additional confidence in the validity of the approach. Much physical insight can be gained by these methods. In the future, it will be useful to use them to discriminate between the various versions of mean-field theories based on slave-particle approaches. A mean-field approach that would contain most of features of numerical approaches might help to gain further insight into the problem. Variational wave functions, even if treated with numerical methods, are also helpful to this end. 
  
Much remains to be done. In the one-band model, it is necessary to develop methods that allow one to reliably investigate long-wavelength instabilities, such as the CDW observed for the cuprates. Investigating theoretically all the details of the highly unusual superconducting properties of the layered organics also remains to be done. And the description of all features of the cuprates at lower temperature calls for more extensive studies of the three-band model.~\cite{Emery:1987,Macridin:2005,Hanke:2010,ScalapinoThread:2010}   
  
Acknowledgments: I am especially grateful to David S\'en\'echal, Giovanni Sordi, Patrick S\'emon, Dominic Bergeron, Yury Vilk, Bumsoo Kyung, Kristjan Haule and to many other collaborators on some of the work discussed above, including V. Bouliane, M. Capone, A. Day, M. Civelli, B. Davoudi, S. Kancharla, G. Kotliar, J.-S. Landry, P.-L. Lavertu, M.-A. Marois, S. Roy. This work was partially supported by the Tier I Canada Research Chair Program (A.-M.S.T.) by CFI, MELS, Calcul Qu\'ebec and Compute Canada. 

%
%




\clearpage

\end{document}